\documentclass[10pt,onecolumn,twoside]{IEEEtran}    

\usepackage{amsmath}

\usepackage{amsfonts}

\usepackage{mathtools}

\usepackage{bbold}

\usepackage{upgreek}

\usepackage{graphicx}

\usepackage{subfigure}

\usepackage{bm}

\usepackage{cite}

\usepackage{color}

\usepackage{CJK}

\usepackage{verbatim}

\usepackage{url}

\usepackage{multirow}

\usepackage{array}

\usepackage{booktabs}

\usepackage{colortbl}

\usepackage{algorithm}

\usepackage{algorithmicx,algpseudocode}

\usepackage{diagbox}

\usepackage{makecell}

\usepackage{enumerate}

\usepackage{cases}

\usepackage{BOONDOX-cal}

\usepackage{varioref} 

\usepackage[misc]{ifsym}

\def\QEDclosed{\mbox{\rule[0pt]{1.3ex}{1.3ex}}} 

\usepackage[thmmarks]{ntheorem}

{
  \theoremstyle{nonumberplain}
  \theoremheaderfont{\bfseries\em}
  \theorembodyfont{\normalfont}
  \theoremseparator{:}
  \theoremsymbol{\mbox{$\QEDclosed$}}
  \newtheorem{Proof}{Proof}
}

\theoremheaderfont{\bf\em}
\theorembodyfont{\normalfont}
\theoremseparator{:}

\newtheorem{Assumption}{Assumption}

\newtheorem{Theorem}{Theorem}

\newtheorem{Lemma}{Lemma}

\newtheorem{Definition}{Definition}

\newtheorem{Remark}{Remark}

\newcommand* \abs[1]{\lvert#1\rvert}

\newcommand* \cb[1]{\bm{#1}}

\begin{document}

\title{False Data Injection Attacks and the Distributed Countermeasure in DC Microgrids} 

\author{Mengxiang Liu, Chengcheng Zhao, Ruilong Deng, Peng Cheng, Wenhai Wang and Jiming Chen
\thanks{A preliminary version of this paper was presented at the IEEE American Control Conference, Philadelphia, PA, July, 2019 \cite{liu2019nonzero}. The authors are with State Key Lab. of Industrial Control Technology, Zhejiang University, Hangzhou, China (e-mails: \{lmx329, chengchengzhao, dengruilong, lunarheart, zdzzlab and cjm\}@zju.edu.cn)}}

\maketitle
\begin{abstract}
In this paper, we consider a hierarchical control based DC microgrid (DCmG) equipped with unknown input observer (UIO) based detectors, where the potential false data injection (FDI) attacks and the distributed countermeasure are investigated. {\color{blue}First, we find that the vulnerability of the UIO-based detector originates from the lacked knowledge of true unknown inputs. Zero trace stealthy (ZTS) attacks can be launched by secretly faking the unknown inputs, under which the detection residual will not be altered, and the impact on the DCmG in terms of voltage balancing and current sharing is theoretically analyzed. Then, to mitigate the ZTS attack, we propose an automatic and timely countermeasure based on the average point of common coupling (PCC) voltage obtained from the dynamic average consensus (DAC) estimator. The integrity of the communicated data utilized in DAC estimators is guaranteed via UIO-based detectors, where the DAC parameters are perturbed in a fixed period to be concealed from attackers. Finally, the detection and mitigation performance of the proposed countermeasure is rigorously investigated, and extensive simulations are conducted in Simulink/PLECS to validate the theoretical results.}
\end{abstract}

\section{Introduction}

During the past decade, the microgrid composed of distributed generation units (DGUs), storage devices, and flexible loads has become one of the most promising solutions to integrate DGUs such as photovoltaic (PV) panels and wind turbines into the power distribution system \cite{hatziargyriou2014microgrids}. In particular, the tremendous growth in DC loads such as laptop computers, LED lights, and telecommunication centers indicates that the DC microgrid (DCmG) would be an economic and feasible solution in addressing the future energy needs \cite{lotfi2017ac}.

In DCmGs, the hierarchical control framework is typically adopted to achieve the voltage balancing and current sharing \cite{meng2017review}. Specifically, the primary control layer regulates the output voltage of the buck converter to track the reference point of common coupling (PCC) voltage. The secondary control layer adjusts the reference PCC voltage by employing centralized or distributed communication networks \cite{dorfler2015breaking}, under which the accuracy of current sharing can be significantly increased. However, the adoption of information and communications technology also brings in new vulnerabilities like the threats of malicious cyberattacks, which could cause economic losses to or even crash the DCmG. Since there exist many special characteristics unique to the DCmG compared with the general cyber-physical system (CPS) like the high interconnectivity, the hierarchical control framework, the flexible network topology \cite{tucci2018stable}, and etc., considerable attention has been attracted to the unique cybersecurity issue therein.

In the power and energy society, the topic of the cybersecurity issue in microgrids has received widespread attention. Considering the microgrid operating in the autonomous mode, Zhang \emph{et al.} \cite{zhang2018distributed} investigated the impact of false data injection (FDI) attacks on distributed load sharing and derived the stable regions under attacks. 
For a well-planned set of {\em balanced} FDI attacks where no physical error is incurred in the DCmG, Sahoo \emph{et al.} \cite{sahoo2018stealth} proposed a cooperative vulnerability factor based anomaly detection framework. In \cite{beg2018signal}, Beg \emph{et al.} proposed a {\em signal temporal logic} based attack detection framework in the DCmG, which can monitor the output voltages and currents against predefined specifications. Nevertheless, the aforementioned literature does not consider the possibility of intelligent attackers, nor proposes the corresponding countermeasure. The intelligent attacker is likely to bypass a certain detector after fully understanding the system model knowledge, and cause specific and accurate adverse effect without being detected. Recent security incidents showed that the intelligent attacker can learn necessary information after penetrating into the system, or collect them from insiders, who have access to critical information legally \cite{chen2010stuxnet}. Hence, it is of great significance to study the possible threats that could be caused by intelligent attacks, and propose the corresponding countermeasure accordingly.

{\color{black}Since the DCmG is a typically CPS, we also review representative literature about the cybersecurity issue in the context of CPSs. In \cite{pasqualetti2013attack}, Pasqualetti \emph{et al.} characterized the undetectable attacks in terms of \emph{zero dynamics}, and designed centralized and distributed attack detection monitors. 
Inspired by the model-based fault diagnosis technique \cite{ding2008model}, Teixeira \emph{et al.} proposed a distributed scheme to detect and isolate cyberattacks utilizing the unknown input observer (UIO), which requires that each agent should have certain global knowledge \cite{teixeira2010networked,teixeira2014distributed}. 
Nevertheless, the aforementioned methods either rely on the centralized entity or require that each agent should have certain global knowledge, which may be not compatible with the scalability property required by the DCmG \cite{tucci2016decentralized}. {\color{blue}Moreover, Barboni \emph{et al.} \cite{9104026} designed a novel distributed observe-based estimation technique for detecting covert attacks, and thoroughly investigated the sufficient detectability conditions. Yet merely local covert attacks inside the subsystem were considered. Recently, Gallo \emph{et al.} \cite{gallo2020distributed} proposed a completely distributed monitoring scheme by combining the Luenberger observers with UIOs, which solely requires the local model knowledge and local information flow, and can be directly applied to the DCmG for the validation of the communicated data between DGUs.} However, it is worth noting that there still exist cyberattacks unforeseeable to the proposed monitoring scheme, and the impact of such attacks has not yet been investigated and mitigated.

Towards this end, in this paper, we investigate the vulnerability of the UIO-based detector, and theoretically analyze the threat of such vulnerability in the context of DCmGs. Furthermore, based on the analysis, we propose an automatic and timely countermeasure against the vulnerability, and its performance in vulnerability perception and threat mitigation is thoroughly studied. {In addition to our preliminary work \cite{liu2019nonzero}, we design a novel distributed countermeasure and provide rigorous theoretical analysis.} {\color{blue}Specifically, the contributions of this paper are listed as follows:
\begin{enumerate}
  \item We find that the vulnerability of the UIO-based detector originates from the lacked knowledge of true unknown inputs. By secretly faking the unknown inputs, the zero trace stealthy (ZTS) attack, which is a special case of the covert attack described in \cite{gallo2020distributed}, can be launched without altering the detection residual. Moreover, we theoretically analyze the impact of both single and cooperative ZTS attacks on the DCmG.
  \item Based on the average PCC voltage (APV) obtained from the dynamic average consensus (DAC) estimator, we propose an automatic and timely countermeasure against ZTS attacks. The DAC parameters are perturbed in a fixed period to be concealed from attackers, such that the integrity of the communicated data utilized in DAC estimators can be guaranteed via UIO-based detectors.
  \item The sufficient condition on detecting ZTS attacks is derived, and the effectiveness of the impact mitigation strategy is rigorously analyzed. Extensive simulations are conducted in Matlab Simulink/PLECS to validate our theoretical results.
\end{enumerate}}

The rest of this paper is organized as follows. Section \ref{section:DCmG model} presents the system model and the problem formulation. Section \ref{section:Undetectable attacks} illustrates the construction of ZTS attacks, and investigates the impact of ZTS attacks in DCmGs. The distributed countermeasure is proposed and elaborated in Section \ref{section:countermeasure}. Finally, simulations results are shown in Section \ref{section:simulation} and Section \ref{section:conclusion and future works} concludes this paper.

\textbf{\emph{Notation:}} $\mathbb{C}$ is the set of complex numbers, and $\mathbb{R}/\mathbb{R}^n$ is the set of real numbers/vectors. The symbol $\abs{\cdot}$ denotes the cardinality of a finite set and component-by-component absolute value of a matrix/vector, and $\|\cdot\|$ represents the norm of a matrix/vector. Inequalities of matrices/vectors are compared component-by-component, and $\lim\limits_{t\to\infty}y(t)$ is denoted by $y(\infty)$ for brevity. Let $\mathbb{1}^n$/$\mathbb{1}^{n\times n}$ and $\mathbb{0}^n$/$\mathbb{0}^{n\times n}$ denote vectors/matrices with all 1 and 0 entries, respectively, and ${\rm I}^n$ denotes the unit matrix with $n\times n$ dimension. Scalar ${v}_{[m]}$ denotes the $m$-th entry of vector $\cb{v}\in\mathbb{R}^n$. Let $\mathbb{H}^1$ denote the subspace of $\mathbb{R}^n$ composed by all vectors satisfying $\langle \cb{v} \rangle=\frac{1}{n}\sum_{i=1}^n{v}_{[i]}=0$, where $\langle \cb{v} \rangle$ denotes the average of all elements in vector $\cb{v}$. Intuitively, each vector in $\mathbb{H}^1$ has $n-1$ freedom\footnote{Any $n-1$ entries in the vector can be set arbitrarily.}, indicating that the dimension of $\mathbb{H}^1$ is $n-1$, i.e., dim$\{\mathbb{H}^1\}=n-1$. Moreover, let $\mathbb{H}^1_\bot$ be the orthogonal subspace of $\mathbb{H}^1$ such that $\mathbb{H}^1 \bigoplus \mathbb{H}^1_\bot = \mathbb{R}^n$, then we have that $\forall \cb{v} \in \mathbb{H}^1_\bot, \cb{v}=\alpha\mathbb{1}^n, \alpha \in \mathbb{R}$ and dim$\{\mathbb{H}^1_\bot\}=1$. Here $\bigoplus$ denotes the direct sum of subspaces.

\vspace{-2pt}
\section{System Model and Problem Formulation}\label{section:DCmG model}

\begin{figure}
  \centering
  \includegraphics[scale=0.25]{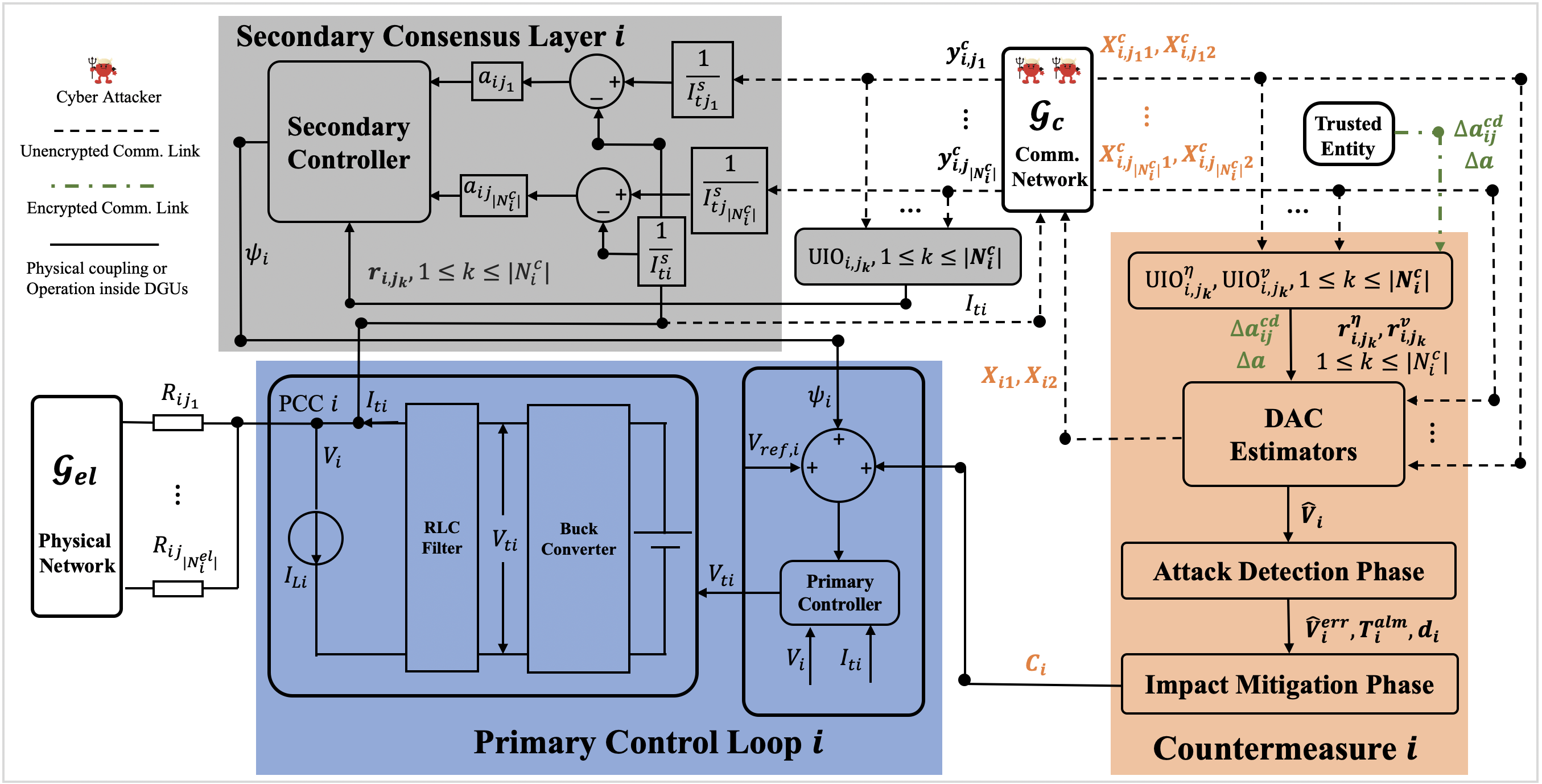}
  \caption{\color{black}{This figure shows the hierarchical control framework and the distributed countermeasure in DGU $i$.}}\label{fig:system_model}
\end{figure}

\subsection{Network Model}
\emph{\textbf{Graph Theory:}} A weighted undirected graph (WUG) is denoted by $\mathcal{G}=\{\mathcal{V}, \mathcal{E}, W\}$, where $\mathcal{V}=\{1,2,\cdots,n\}$ is the set of nodes, $\mathcal{E} = \{(i,j)\} \subseteq \mathcal{V} \times \mathcal{V}$ is the set of edges, and $W={\rm diag}\{a_{ij}\}\in\mathbb{R}^{|\mathcal{E}| \times |\mathcal{E}|}$ is the diagonal matrix composed by edge weights $a_{ij}, \forall (i,j)\in\mathcal{E}$. The set of neighbors of node $i$ is $\mathcal{N}_i = \{j|(i,j) \in \mathcal{E} \}$. After assigning each edge of $\mathcal{G}$ an arbitrary direction, the oriented incidence matrix is computed as $P(\mathcal{G})\in \mathbb{R}^{|\mathcal{V}|\times|\mathcal{E}|}$, where the order of columns corresponds to the order of edge weights in $W$ \cite{francesco2018lectures}. Then, the Laplacian matrix of $\mathcal{G}$ can be expressed as $\mathcal{L}(\mathcal{G})=P(\mathcal{G})WP(\mathcal{G})^{\rm T}$, which is independent of the edge orientations.

\emph{\textbf{Electrical and Communication Networks:}} The electrical network of DCmG is represented by WUG $\mathcal{G}_{el}=\{\mathcal{V},\mathcal{E}_{el},W_{el}\}$, where nodes are DGUs, edges are power lines whose orientations define reference directions for positive currents. Moreover, edge weights are conductances of power lines. The set of neighbors of node $i$ is $\mathcal{N}_i^{el}$, $|\mathcal{V}|=N$ and the Laplacian matrix is $\mathcal{L}(\mathcal{G}_{el})=M$. The communication network of DCmG is denoted by WUG $\mathcal{G}_c=\{\mathcal{V},\mathcal{E}_{c},W_c\}$, where edges are communication links and edge weights are $a_{ij}^c, \forall (i,j) \in \mathcal{E}_{c}$. The set of neighbors of node $i$ is $\mathcal{N}_i^c$, and the Laplacian matrix is $\mathcal{L}(\mathcal{G}_c)=L$.

\subsection{DGU Dynamics}

As shown in Fig. \ref{fig:system_model}, each DGU contains a DC voltage source, a buck converter, a local load current, and a RLC (resistor, inductor, and capacitor) filter. Notice that $V_{ti}$ is the output voltage of DC buck converter and $I_{Li}$ is the constant load current. Moreover, $V_i$ and $I_{ti}$ are the PCC voltage and the output current, respectively. The hierarchical control framework is deployed in each DGU, where the primary controller tracks the local reference PCC voltage and the secondary consensus layer regulates the local reference PCC voltage to achieve current sharing and voltage balancing in the DCmG \cite{tucci2018stable}. The dynamical model of DGU $i \in \mathcal{V}$ is
{\begin{align}\label{eq:state space model}
\left\{
\begin{aligned}
\cb{\dot{x}}_i(t)  =\ & A_{ii}\cb{x}_i(t)+\cb{b}_iu_i(t)+\cb{g}_i\psi_i(t)+M_i\cb{d}_i+\cb{\xi}_i(t)+\cb{\omega}_i(t)\\
\cb{y}_i(t) =\ & \cb{x}_i(t)+\cb{\rho}_i(t) 
\end{aligned}
\right.,
\end{align}}where $\cb{x}_i(t)=[V_i(t),I_{ti}(t),v_i(t)]^{\rm{T}}$ is the state vector, and $v_i(t)$ is the integral of the voltage tracking error defined by $\dot{v}_i(t)=V_{ref,i}+\psi_i(t)-V_i(t)$. Here $V_{ref,i}$ is the nominal reference PCC voltage and $\psi_i(t)$ is the secondary control input. Moreover, $\cb{d}_i=[I_{Li},V_{ref,i}]^{\rm{T}}$ is the constant exogenous input vector, and $\cb{y}_i(t)\in\mathbb{R}^3$ is the output vector. The physical couplings with neighboring DGUs are modeled as $\cb{\xi}_i(t)=\sum_{j \in \mathcal{N}_i^{el}} A_{ij}\cb{x}_j(t) \in \mathbb{R}^3$. 
The primary control input is computed as
{\begin{align}\label{eq:primary control input}
u_i(t)=V_{ti}=\cb{k}_i^{\rm{T}}\cb{y}_i(t),
\end{align}}where the primary control gain $\cb{k}_i \in \mathbb{R}^3$ depends merely on the model knowledge of DGU $i$ and the interconnected power lines \cite{tucci2016decentralized}. The secondary control input is obtained through the following consensus scheme, i.e.,
{\begin{align}{\label{eq:secondary control input}}
\dot{\psi_i}(t)=-[0,k_{I},0]\sum_{j \in \mathcal{N}_i^c}a_{ij}^c(\frac{\cb{y}_{i}(t)}{I_{ti}^{s}}-\frac{\cb{y}_{i,j}^c(t)}{I_{tj}^{s}}),
\end{align}}where $\cb{y}_{i,j}^c(t)$ is the output of DGU $j$ communicated to DGU $i$, $I_{ti}^s>0$ and $I_{tj}^s>0$ are rated currents corresponding to DGU $i$ and DGU $j$, respectively, and $k_I>0$ is the weight parameter invariant among all DGUs. We have the following Assumptions regarding to the DCmG model.

\begin{Assumption}\label{Ass:noise}
The process noise and measurement noise are unknown-but-bounded (UBB) i.e., $\abs{\cb{\omega}_i(t)} \leq \bar{\cb{\omega}}_i \in \mathbb{R}^3, \abs{\cb{\rho}_i(t)} \leq \bar{\cb{\rho}}_i \in \mathbb{R}^3,\forall t \geq 0$.
\end{Assumption}

\begin{Assumption}{\label{Ass:Voltage}}
The nominal reference PCC voltages are equal among all DGUs, i.e., $V_{ref,i}=V_{ref}, \forall i \in \mathcal{V}$.
\end{Assumption}

\begin{Assumption}{\label{Ass:graph}}
The WUGs $\mathcal{G}_{c}$ and $\mathcal{G}_{el}$ are both connected, and they have the same topology and edge weights, i.e., $L=M$.
\end{Assumption}


Under Assumptions \ref{Ass:noise}-\ref{Ass:graph}, the hierarchical control framework \eqref{eq:primary control input}-\eqref{eq:secondary control input} can achieve voltage balancing and current sharing in DCmGs \cite{tucci2018stable}, which are formally defined as
\begin{Definition}[Voltage Balancing]\label{def:voltage balancing}
Under Assumption \ref{Ass:Voltage}, voltage balancing is achieved if $\langle\cb{v}(\infty)\rangle=V_{ref}$, where $\cb{v}(t)=[V_i(t),\cdots,V_N(t)]^{\rm T}$, and $\langle \cb{v}(\infty) \rangle$ denotes the steady-state APV.
\end{Definition}
\begin{Definition}[Current Sharing]\label{def:current sharing}
For constant load currents, current sharing is achieved if $\frac{I_{ti}(\infty)}{I_{ti}^s}=\frac{I_{tj}(\infty)}{I_{tj}^s}$, $\forall i,j \in \mathcal{V}$, i.e., load currents are shared proportionally to the rated currents.
\end{Definition}

\subsection{Attack Model}
In this paper, we consider FDI attacks injecting malicious signals into communication links between DGUs. In particular, the FDI attack against $(i,j)\in \mathcal{E}_c$ is modeled as
\begin{align}\label{eq:attack model}
\cb{y}_{i,j}^c(t)=\cb{y}_j(t)+\beta(t-T_a)\cb{\phi}_{i,j}(t),
\end{align}where $\cb{\phi}_{i,j}(t)$ is an arbitrary vector designed by the attacker, and $\beta(t-T_a)$ is the step function with $T_a$ delay. The attack is started at $t=T_a$, i.e., $\cb{y}_{i,j}^c(t)=\cb{y}_j(t), \forall t \le T_a$. {\color{blue}In this study, we consider the continuous and differentiable attack vector and give the following assumption.
\begin{Assumption}\label{Ass:Attack Vector}
The attack vector $\cb{\phi}_{i,j}(t)$ is continuous and differentiable.
\end{Assumption}
\begin{Remark}
Assumption \ref{Ass:Attack Vector} is practical as it can guarantee the smoothness of the corrupted signal, such that the corrupted signal would be indistinguishable from the normal signal. Moreover, the resulting conclusion under Assumption \ref{Ass:Attack Vector} would have explicit forms and could further facilitate our future research on more general attack vectors.
\end{Remark}}

{\color{black}Actually, the attacker is likely to obtain system parameters from the insider \cite{Hunker2011InsidersAI}, who can get access to them legally. But the attacker is hard to obtain real-time system parameters as the insider only discloses them to the attacker in a specific time period to guarantee his/her hiddenness. Therefore, we consider that the attacker is able to obtain system parameters involved in DGU dynamics \eqref{eq:state space model} every few hours or days (not immediately). Moreover, the attacker is able to eavesdrop the communicated data through IP spoofing attacks. Nevertheless, the attacker cannot intrude into DGU $i$ or the DCmG control center\footnote{In the DCmG, the control center is mainly responsible for the tertiary control layer including optimal operation in grid-tied and islanded operating modes and power flow control in grid-tied mode \cite{bidram2012hierarchical}.} due to various host-based defense mechanisms \cite{crosbie1995defending}, indicating that the primary control input $u_i(t)$ will not be compromised.}


\vspace{-10pt}
\subsection{UIO-based Detector}
According to \cite{gallo2020distributed}, a bank of UIOs are deployed in each DGU to identify and isolate FDI attacks \eqref{eq:attack model} among the neighboring communication links. First, the dynamical model of DGU $j \in \mathcal{N}_i^c$ \eqref{eq:state space model} is transformed to
{\begin{align}\label{eq:typical UIO}
\left\{
\begin{aligned}
\dot{\cb{x}}_j(t)&=A_{kj}\cb{x}_j(t)+\bar{E}_j\bar{\cb{d}}_j(t)+\cb{\omega}_j(t)+\cb{b}_j\cb{k}_j^{\rm T}\cb{\rho}_j(t)\\
\cb{y}_j(t)&=\cb{x}_j(t)+\cb{\rho}_j(t)
\end{aligned}
\right.,
\end{align}}where $A_{kj}=A_{jj}+\cb{b}_j\cb{k}_j^{\rm{T}} \in \mathbb{R}^{3\times3}$, $\bar{E}_j\bar{\cb{d}}_j(t) = M_j\cb{d}_j+\cb{g}_j\psi_j(t)+\cb{\xi}_j(t)$, and $\bar{E}_j\in \mathbb{R}^{3\times2}$ is a full column rank matrix related to the capacitor parameter $C_{tj}$ as shown in \eqref{E_j&H_j}.
{\begin{align}\label{E_j&H_j}
\bar{E}_j=\left[ 
\begin{array}{ccc}
\frac{1}{C_{tj}} & 0 & 0\\
0   & 0 & 1
\end{array} 
\right]^{\rm{T}},
H_j=\left[ \begin{array}{ccc}
1 & 0  & 0 \\
h_{12} & h_{22} & h_{32} \\
0 &  0 & 1
\end{array}
\right]^{\rm{T}}.
\end{align}}Moreover, vector $\bar{\cb{d}}_j(t)$ contains the inputs of DGU $j$ unknown to DGU $i$. Based on \eqref{eq:typical UIO}, one can easily verify that ${\rm rank}({\rm I}^3\bar{E}_j)={\rm rank}(\bar{E}_j)$ and matrix 
$\left[ 
\begin{array}{lll}
s{\rm I}^3-A_{kj} & \bar{E}_j\\
{\rm I}^3   & \mathbb{0}^{3\times 2}
\end{array} 
\right]$ has full column rank $\forall s \in \mathbb{C}$. Hence, according to Theorem 1 in \cite{chen1996design}, the full order UIO in DGU $i$ can be constructed as
{\begin{align}\label{eq:UIO}
{\rm UIO}_{i,j} \left \{
\begin{aligned}
&\dot{\cb{z}}_{i,j}(t)=F_j\cb{z}_{i,j}(t)+\hat{K}_j\cb{y}_{i,j}^c(t)\\
&\hat{\cb{x}}_{i,j}(t)=\cb{z}_{i,j}(t)+H_j\cb{y}_{i,j}^c(t)
\end{aligned}
\right.,
\end{align}}under which, in the normal case, the estimated state $\hat{\cb{x}}_{i,j}(t)$ will converge asymptotically to $\cb{x}_j(t)$ regardless of the unknown input vector $\bar{\cb{d}}_j(t)$. Here $\cb{z}_{i,j}(t) \in \mathbb{R}^3$ is the internal state of UIO \eqref{eq:UIO}, and UIO parameters $F_j, \hat{K}_j, H_j \in \mathbb{R}^{3\times3}$ need to satisfy
{\begin{subequations}\label{eq:UIO parameters}
\begin{alignat}{2}
&T_j\bar{E}_j=\mathbb{0}^{3\times2},\label{eq:UIO parameters 3}\\
&T_j={\rm I}^3-H_j,\label{eq:UIO parameters 2} \\
&\hat{K}_j=K_{j1}+K_{j2},\label{eq:UIO parameters 1}\\
&F_j=T_jA_{kj}-K_{j1},\label{eq:UIO parameters 4}\\
&K_{j2}=F_jH_j,\label{eq:UIO parameters 5}
\end{alignat}
\end{subequations}}where $K_{j1},K_{j2},T_j \in \mathbb{R}^{3\times3}$, $H_j$ is defined in \eqref{E_j&H_j}, $h_{12}, h_{22}, h_{32}$ are arbitrary scalars, and $K_{j1}$ should be appropriately chosen to make the eigenvalues of $F_j$ all lie in the open left half-plane based on \eqref{eq:UIO parameters 4}. In the absence of attacks, the analytical expression of detection residual $\cb{r}_{i,j}(t)=\cb{y}_{i,j}^c(t)-\hat{\cb{x}}_{i,j}(t)$ can be obtained given DGU dynamics \eqref{eq:typical UIO} and UIO \eqref{eq:UIO}, i.e., 
\begin{align}\label{eq:expression of residual}
\cb{r}_{i,j}(t)=e^{F_jt}(\cb{\sigma}_{2i,j}(0)+\cb{\sigma}_{3i,j}(t))+T_j\cb{\rho}_j(t),
\end{align}where $\cb{\sigma}_{2i,j}(0)=\cb{x}_j(0)-\hat{\cb{x}}_{i,j}(0)+H_j\cb{\rho}_j(0)$ and $\cb{\sigma}_{3i,j}(t)=\int_0^te^{-F_j\tau}(T_j\cb{\omega}_j(\tau)+(T_j\cb{b}_j\cb{k}_j-\hat{K}_j)\cb{\rho}_j(\tau)){\rm d}\tau$. Since $\cb{y}_j(0)=\cb{x}_j(0)+\cb{\rho}_j(0)$, $\cb{z}_{i,j}(0)$ can be set as $T_j\cb{y}_{i,j}^c(0)$ such that the initial state estimation error is bounded by the bound of measurement noise in the absence of attacks, i.e., 
{\begin{align*}
    |\cb{x}_j(0)-\hat{\cb{x}}_{i,j}(0)|=|\cb{y}_j(0)-\cb{y}_{i,j}^c(0)-\cb{\rho}_j(0)| = |\cb{\rho}_j(0)| \le \bar{\cb{\rho}}_j.
\end{align*}}Moreover, as $F_j$ is Hurwitz stable, there exist positive constants $\kappa, \mu$ such that $||e^{F_jt}|| \le \kappa e^{-\mu t},\forall t \ge 0$. Then, the time-varying detection threshold $\cb{\bar{r}}_{i,j}(t)$ is computed such that
{\begin{align}\label{eq:normal condition 2}
|\cb{r}_{i,j}(t)|\le\cb{\bar{r}}_{i,j}(t)=\kappa e^{-\mu t}(\bar{\cb{\sigma}}_{2i,j}(0)+\bar{\cb{\sigma}}_{3i,j}(t))+|T_j|\bar{\cb{\rho}}_j
\end{align}}always hold in the absence of attacks, where $|\cb{\sigma}_{2i,j}(0)|\le\bar{\cb{\sigma}}_{2i,j}(0)=({\rm I}^3+|H_j|)\bar{\cb{\rho}}_j$ and $|\cb{\sigma}_{3i,j}(t)|\le\bar{\cb{\sigma}}_{3i,j}(t)=\int_0^t|e^{-F_j\tau}|(|T_j|\bar{\cb{\omega}}_j+|T_j\cb{b}_j\cb{k}_j-\hat{K}_j|\bar{\cb{\rho}}_j){\rm d}\tau$. 

Once \eqref{eq:normal condition 2} is violated, it is considered that the data $\cb{y}_{i,j}^c(t)$ received from DGU $j$ is corrupted by attacks. With some abuse of the notation, let $\cb{r}_{i,j}(t)$ be the detection residual under attacks, and it is decomposed as 
\begin{align*}
\cb{r}_{i,j}(t)=\tilde{\cb{r}}_{i,j}(t)+\cb{r}_{i,j}^a(t),
\end{align*}where $\tilde{\cb{r}}_{i,j}(t)$ is the healthy residual component equating to \eqref{eq:expression of residual} and $\cb{r}_{i,j}^a(t)$ is the malicious component associated with attacks. Given the attack model \eqref{eq:attack model}, DGU dynamics \eqref{eq:typical UIO} and UIO \eqref{eq:UIO}, we obtain 
{\begin{align}\label{eq:residuals impacts}
\begin{aligned}
\cb{r}_{i,j}^a(t)=e^{F_j(t-T_a)}H_j\cb{\phi}_{i,j}(T_a)+T_j\cb{\phi}_{i,j}(t)-\int_{T_a}^te^{F_j(t-\tau)}\hat{K}_j\cb{\phi}_{i,j}(\tau){\rm d}\tau.
\end{aligned}
\end{align}}



\vspace{-10pt}
\subsection{Problems of Interest}
{\color{blue}In this paper, we are interested in the FDI attacks that cause no impact on the detection residual, i.e., $\cb{r}_{i,j}^a(t)=\mathbb{0}^3$, while the received output $\cb{y}_{i,j}^c(t)$ deviates a lot from the true output $\cb{y}_{j}(t)$.} For clarity, we define the FDI attack aforementioned as
\begin{Definition}[ZTS Attack]\label{Def:ZTS}
Given DGU dynamics \eqref{eq:typical UIO} and UIO-based detector \eqref{eq:UIO}, the FDI attack \eqref{eq:attack model} is \textbf{ZTS} if
{\begin{align*}
\left\{
\begin{aligned}
&\cb{\phi}_{i,j}(t) \ne \mathbb{0}^3, \exists t \ge 0\\
&\cb{r}_{i,j}^a(t)= \mathbb{0}^3, \forall t \ge 0
\end{aligned}
\right..
\end{align*}}
\end{Definition}

\begin{Remark}
{\color{blue}Zero-dynamics attacks characterize a class of undetectable attacks that excite only \emph{zero dynamics} of a dynamical system, which can make the system states diverge while leave no trace on the outputs, and thus are inherently undetectable for detectors. According to the attack model in \cite{pasqualetti2013attack}, the ZTS attack corrupting the outputs of a dynamical system can be described by the Rosenbrock matrix $P(s)=\left[ \begin{array}{cc}
s{\rm I}^3-A_{kj} & \mathbb{0}^{3\times3} \\
{\rm I}^3 & {\rm I}^3
\end{array}
\right]$, under which the \emph{zero dynamics} of the system can never be excited with $B=\mathbb{0}^{3\times3}$. Hence, ZTS attacks are essentially different from zero-dynamics attacks. Specifically, ZTS attacks reveal the vulnerability of the attack detection and identification for a dynamical system when there exist some \emph{unknown inputs} regardless of the \emph{zero dynamics}.}
\end{Remark}

The following three problems are formulated: (1) How can the attacker construct ZTS attacks? \textbf{(P1)} (2) How will ZTS attacks affect the DCmG? \textbf{(P2)} (3) How to detect and mitigate the impact of ZTS attacks? \textbf{(P3)}

\section{ZTS Attack and the Impact Analysis}\label{section:Undetectable attacks}
In this section, we characterize the condition for the FDI attack \eqref{eq:attack model} to be ZTS and investigate the impact of ZTS attacks in DCmGs.
\subsection{ZTS Attack}
{\color{blue}Although DGU $i$ can estimate (recover) the unknown inputs of DGU $j\in\mathcal{N}_i$ from $\cb{y}_{i,j}^c(t)$, it is still not sure whether the estimated (recovered) unknown inputs are \emph{true} or not. Hence, the intuition is to deceive DGU $i$ utilizing a fake unknown input vector $\cb{d}_{ij}^a(t)$, which motivates the following analysis.}
\begin{Theorem}\label{Theo:ZTS attack}
Given DGU dynamics \eqref{eq:typical UIO} and UIO-based detector \eqref{eq:UIO}, {\color{blue}the FDI attack \eqref{eq:attack model} under Assumption \ref{Ass:Attack Vector}} is ZTS if and only if the attack vector $\cb{\phi}_{i,j}(t)$ satisfies
{\begin{align}\label{eq:dynamics of ZTS}
\left\{
  \begin{aligned}
    & \dot{\cb{\phi}}_{i,j}(t)=A_{kj}\cb{\phi}_{i,j}(t)+\bar{E}_j\bar{\cb{d}}_{ij}^{a}(t),\forall t\ge T_a \\ 
    &\cb{\phi}_{i,j}(T_a)=\mathbb{0}^3
  \end{aligned}
\right.,
\end{align}}where $\bar{\cb{d}}_{ij}^{a}(t)$ should be designed such that $\bar{E}_j\bar{\cb{d}}_{ij}^{a}(t)\neq\mathbb{0}^3$.
\end{Theorem}
\begin{Proof}
{\em(If)} According to the DGU dynamics \eqref{eq:typical UIO}, the Laplace form of $\cb{y}_{i,j}^c(t)$ corrupted by the attack vector $\cb{\phi}_{i,j}(t)$ satisfying \eqref{eq:dynamics of ZTS} is 
{\begin{align}\label{Corrupted measuremnet}
\cb{y}_{i,j}^c(s)=(s{\rm I}^3-A_{kj})^{-1}(\cb{x}_j(0)+\bar{E}_j\tilde{\cb{d}}_j(s)+\cb{\omega}_j(s)+\cb{b}_j\cb{k}_j\cb{\rho}_j(s))+\cb{\rho}_j(s),
\end{align}}where $\tilde{\cb{d}}_j(s)=\bar{\cb{d}}_j(s)+\bar{\cb{d}}_{ij}^{a}(s)$ integrates the normal and fake unknown input vectors. It follows from \eqref{Corrupted measuremnet} that $\cb{y}_{i,j}^c(t)$ can be interpreted as the output of system \eqref{eq:typical UIO} whose unknown input vector $\bar{\cb{d}}_j(t)$ is switched to $\tilde{\cb{d}}_j(t)$ at $t=T_a$. Hence, the attack vector $\cb{\phi}_{i,j}(t)$ satisfying \eqref{eq:dynamics of ZTS} will not alter the detection residual $\cb{r}_{i,j}(t)$, as $\cb{r}_{i,j}(t)$ is designed to be insensitive to the variation of unknown inputs. The proof of the sufficient part is completed.

{\em (Only If)} Suppose that there exists the FDI attack \eqref{eq:attack model} such that $\cb{r}_{i,j}^a(t)= \mathbb{0}^3,\forall t \ge 0$, {\color{blue}then under Assumption \ref{Ass:Attack Vector}, we obtain}
{\begin{align}\label{Solution of ZTS}
\cb{r}_{i,j}^a(T_a)=\mathbb{0}^3, \dot{\cb{r}}_{i,j}^a(t)=\mathbb{0}^3, \forall t \ge T_a.
\end{align}}Substituting \eqref{eq:residuals impacts} into $\cb{r}_{i,j}^a(T_a)=\mathbb{0}^3$, one obtains $(H_j+T_j)\cb{\phi}_{i,j}(T_a)=\mathbb{0}^3$. Then, it follows from \eqref{eq:UIO parameters 2} that $\cb{\phi}_{i,j}(T_a)=\mathbb{0}^3$. Moreover, substituting $\cb{\phi}_{i,j}(T_a)=\mathbb{0}^3$ into \eqref{eq:residuals impacts}, one can transform $\dot{\cb{r}}_{i,j}^a(t)=\mathbb{0}^3,\forall t \ge T_a$ into
{\begin{align}\label{eq:14}
T_j\dot{\cb{\phi}}_{i,j}(t)=(F_jT_j+\hat{K}_j)\cb{\phi}_{i,j}(t),\forall t \ge T_a.
\end{align}}Based on equations \eqref{eq:UIO parameters 2}-\eqref{eq:UIO parameters 5}, we have
{\begin{align}\label{eq:30}
F_jT_j+\hat{K}_j=F_j({\rm I}^3-H_j)+\hat{K}_j=F_j+K_{j1}=T_jA_{kj}.
\end{align}}Substituting \eqref{eq:30} into \eqref{eq:14}, we obtain
{\begin{align}\label{eq:16}
T_j(\dot{\cb{\phi}}_{i,j}(t)-A_{kj}\cb{\phi}_{i,j}(t))=\mathbb{0}^3,\forall t \ge T_a.
\end{align}}Furthermore, it follows from equations \eqref{E_j&H_j}, and \eqref{eq:UIO parameters 3}-\eqref{eq:UIO parameters 2} that $T_j\bar{E}_j=\mathbb{0}^{3\times2}$ and ${\rm rank}(T_j)+{\rm rank}(\bar{E}_j)=3$, indicating that the null space of $T_j$ coincides with the range space of $\bar{E}_j$. Thus, \eqref{eq:16} is equivalent to 
\begin{align}\label{eq:15}
\dot{\cb{\phi}}_{i,j}(t)=A_{kj}\cb{\phi}_{i,j}(t)+\bar{E}_j\bar{\cb{d}}_{ij}^{a}(t),\forall t \ge T_a,
\end{align}under which $\cb{\phi}_{i,j}(t)$ will not be constant zero once $\bar{E}_j\bar{\cb{d}}_{ij}^{a}(t)\neq\mathbb{0}^3$. The proof of the necessary part is completed.
\end{Proof}

Based on \eqref{eq:dynamics of ZTS}, the attacker can construct ZTS attacks once she/he could get access to $A_{kj}, \bar{E}_j$, which are determined by electrical parameters (resistance, capacitance, and inductance) and the primary control gain ($\cb{k}_j$).

\begin{Remark}\label{fundamental limitation}
{\color{blue}Under Assumption \ref{Ass:Attack Vector}, ZTS attacks can only be constructed by utilizing the fake unknown input $\cb{d}_{ij}^a(t)$, indicating that the vulnerability of the UIO-based detector \eqref{eq:UIO} originates from the lacked knowledge of true unknown inputs.} Moreover, as shown in \cite{liu2019nonzero}, the attacker can also construct the stealthy FDI attack \eqref{eq:attack model} (different from the ZTS attack \eqref{eq:dynamics of ZTS}) by utilizing the system noises, under which the detection residual $\cb{r}_{i,j}(t)$ is affected but still bounded by the detection threshold $\bar{\cb{r}}_{i,j}(t)$. Nevertheless, the attacker needs to obtain the knowledge of system noise related terms $\cb{\sigma}_{2i,j}(0),\cb{\sigma}_{3i,j}(T_a)$, which may be impractical. Further, the impact caused by that stealthy FDI attack is limited by bounds of system noises.
\end{Remark}

\begin{Remark}
{\color{blue}
Indeed, the ZTS attack is a special case of the covert attack described in \cite{gallo2020distributed}, with the attack vector being initialized at zero. Under the specific initial condition, we have the sufficient and necessary condition for the attack satisfying Assumption 4 to be ZTS. Moreover, it is noted that some ZTS attacks whose attack vectors are either discontinuous or non-differentiable also exist, but the investigation on them is still challenging due to the diverse and non-unified attack vector forms, and is left as our future work.}
\end{Remark}

\subsection{Attack Impact Analysis} \label{section: Attack impacts}
In this section, we theoretically analyze the impact of ZTS attacks on voltage balancing and current sharing. According to \eqref{eq:dynamics of ZTS}, the pair $(A_{kj},\bar{E}_j)$ is controllable, and thus the ZTS attack vector $\cb{\phi}_{i,j}(t)$ can be arbitrarily large with well-designed $\bar{\cb{d}}_{ij}^a(t)$. However, $\cb{\phi}_{i,j}(t)$ should be bounded to make the corrupted measurement $\cb{y}_{i,j}^c(t)$ physically reachable, given the maximal/minimal PCC voltage and output current for DGU $j$. Hence, we provide the assumption below.
\begin{Assumption}\label{Ass:consatnt faked unknown inputs}
The fake unknown input vector involved in \eqref{eq:dynamics of ZTS}, i.e., $\cb{d}_{ij}^a(t)$, is a bounded constant vector.
\end{Assumption}
\begin{Remark}
Since $A_{kj}$ is Hurwitz stable, the attack vector generated by \eqref{eq:dynamics of ZTS} with constant $\cb{d}_{ij}^a(t)$ will eventually converge. Hence, from the perspective of the attacker, it is also practical and useful to set $\cb{d}_{ij}^a(t)$ as a constant vector, under which bounded $\cb{\phi}_{i,j}(t)$ could be generated at his/her will. Moreover, the impact of the ZTS attacks with time-varying $\cb{d}_{ij}^a(t)$ can be analyzed in a similar way referring to the following results.
\end{Remark}

In the remainder of this paper, $\bar{\cb{d}}_{ij}^a$ is utilized to denote the constant vector $\bar{\cb{d}}_{ij}^a(t)$. With some abuse of notations, let $\cb{\psi}(t)=[\psi_1(t),\cdots,\psi_N(t)]$ be the secondary control input vector  under attacks. Similar to the detection residual $\cb{r}_{i,j}(t)$, $\cb{\psi}(t)$ is  decomposed as $\cb{\psi}(t)=\tilde{\cb{\psi}}(t)+\cb{\psi}_a(t)$, where $\tilde{\cb{\psi}}(t)$ denotes the healthy component and $\cb{\psi}_a(t)$ is the malicious component associated with attacks.
\begin{Theorem}\label{theorem:attack impacts}
Under Assumptions \ref{Ass:Voltage}-\ref{Ass:consatnt faked unknown inputs}, any single ZTS attack \eqref{eq:dynamics of ZTS} will cause
{\begin{align}\label{eq:single attack impacts}
\langle \cb{\psi}_a(\infty) \rangle=-\frac{k_Ia_{ij}^c}{NI_{tj}^s}\cb{k}^{\rm T}A_{kj}^{-1}(\bar{E}_j\bar{\cb{d}}_{ij}^a*(t-T_a)+A_{kj}^{-1}\bar{E}_j\bar{\cb{d}}_{ij}^a),
\end{align}}where $\cb{k}^{\rm T}=[0,1,0]$. Intuitively, with nonzero $\cb{k}^{\rm T}A_{kj}^{-1}\bar{E}_j\bar{\cb{d}}_{ij}^a$, neither voltage balancing nor current sharing can be achieved in DCmGs.
\begin{Proof}
{\color{black}The proof is given in Appendix \ref{appendix:theorem 2}.}
\end{Proof}
\end{Theorem}

Next, we consider the case where ZTS attacks \eqref{eq:dynamics of ZTS} are injected into multi communication links $\tilde{\mathcal{E}}_{c} \subseteq \mathcal{E}_c$ cooperatively such that 
{\begin{align}\label{eq:cooperative attacks conditions}
\sum_{(i,j) \in \tilde{\mathcal{E}}_{c}}\frac{k_Ia_{ij}^c}{I_{tj}^s}\cb{k}^{\rm T}A_{kj}^{-1}\bar{E}_j\bar{\cb{d}}_{ij}^a=0.
\end{align}}
\begin{Theorem}\label{Theorem:Cooperative attack imapcts}
Under Assumptions \ref{Ass:Voltage}-\ref{Ass:consatnt faked unknown inputs}, the cooperative ZTS attacks \eqref{eq:dynamics of ZTS} satisfying \eqref{eq:cooperative attacks conditions} will cause
{\begin{align}\label{eq:cooperative attacks impacts}
\langle \cb{\psi}_a(\infty) \rangle=-\sum_{(i,j) \in \tilde{\mathcal{E}}_{c}}\frac{k_Ia_{ij}^c}{NI_{tj}^s}\cb{k}^{\rm T}A_{kj}^{-2}\bar{E}_j\bar{\cb{d}}_{ij}^a,
\end{align}}under which voltage balancing cannot be achieved if $\sum_{(i,j) \in \tilde{\mathcal{E}}_{c}} \frac{k_Ia_{ij}^c}{I_{tj}^s}\cb{k}^{\rm T}A_{kj}^{-2}\bar{E}_j\bar{\cb{d}}_{ij}^a\ne 0$ and current sharing cannot be achieved if $\sum_{(i,j) \in \tilde{\mathcal{E}}_{c}}\frac{k_Ia_{ij}^c}{I_{tj}^s}\cb{k}^{\rm T}A_{kj}^{-1}\bar{E}_j\bar{\cb{d}}_{ij}^a\cb{l}_{i}\ne\mathbb{0}^N$.
\end{Theorem}
\begin{Proof}
{\color{black}The proof is given in Appendix \ref{appendix:Theorem 3}.}
\end{Proof}
\begin{Remark}
From the perspective of the attacker, he/she can choose appropriate attack vectors $\cb{\phi}_{i,j}(t), \forall (i,j) \in \tilde{\mathcal{E}}_{c}$ referring to the theoretical results in Theorems \ref{theorem:attack impacts}-\ref{Theorem:Cooperative attack imapcts} to achieve his/her malicious goals. Specifically, if the attacker can get access to any communication link $(i,j)\in \mathcal{E}_{c}$, then the single ZTS attack \eqref{eq:dynamics of ZTS} with nonzero $\cb{k}^{\rm T}A_{kj}^{-1}\bar{E}_j\bar{\cb{d}}_{ij}^a$ can destabilize the DCmG. Moreover, if the attacker can get access to multi communication links $\tilde{\mathcal{E}}_{c}$ simultaneously, then he/she can launch the cooperative ZTS attacks \eqref{eq:dynamics of ZTS} satisfying \eqref{eq:cooperative attacks conditions} to induce accurate and specific adverse impact on voltages and currents.
\end{Remark}

\section{The Distributed Countermeasure}\label{section:countermeasure}

In this section, we propose an automatic and timely countermeasure against ZTS attacks based on the APV obtained from the DAC estimator. As shown in Algorithm \ref{Algorithm: the whoel countermeasure}, the countermeasure is composed of two phases, i.e., attack detection and impact mitigation. In particular, the former phase is to reveal the existence of ZTS attacks by utilizing the detection indicator derived from the APV. Once the detection indicator exceeds a predefined threshold, the latter phase is activated for compensation until the voltage balancing is recovered. In the following subsections, we will introduce the DAC estimator, the attack detection phase, and the impact mitigation phase.

\begin{algorithm}[htb]
\caption{The Distributed Countermeasure in DGU $i\in \mathcal{V}$}\label{Algorithm: the whoel countermeasure}
\begin{algorithmic}[1]
\Require{
The PCC voltage $V_i(t)$;
}
\State{Deploy the DAC estimator \eqref{eq:DAC estimator} satisfying \eqref{eq:minimal realizations of LTIs};}
\Ensure{\textbf{\em Attack Detection Phase}}
\State Calculate the detection indicator $\mathcal{d}_i(t)$ according to \eqref{Calculation d_i};
\State Set the detection threshold $\bar{\mathcal{d}}_i$ according to \eqref{eq:detection threshold};
\If{$\mathcal{d}_i(t)>\bar{\mathcal{d}}_i$}
\State Activate the impact mitigation phase;
\Else{}
\State Repeat the detection phase;
\EndIf
\Ensure{\textbf{\em Impact Mitigation Phase}}
\State Compute the compensation value $C_i(t)$ according to \eqref{eq:Compensator};
\State Add $C_i(t)$ to the secondary control input $\psi_i(t)$ according to \eqref{eq:impact counteraction};
\If{Condition \eqref{eq:compensation completed condition} is satisfied} \Comment{{ \em  Judge whether voltage balancing is recovered}}
\State Goto the detection phase;
\Else{}
\State Repeat the impact mitigation phase;
\EndIf
\end{algorithmic}
\end{algorithm}

\vspace{-15pt}
\subsection{The DAC Estimator}
In this subsection, we introduce the DAC estimator equipped with UIO-based detectors, which are utilized to validate the integrity of the communicated data between DGUs. From \cite{bai2010robust}, the dynamics of the DAC estimator in DGU $i\in \mathcal{V}$ follow
{ \begin{subequations}\label{eq:DAC estimator}
\begin{alignat}{2}
&\left\{
  \begin{aligned}
    &\dot{\cb{X}}_{i1}(t)=A_1\cb{X}_{i1}(t)+B_1(V_i(t)-\gamma\sum_{j \in \mathcal{N}_i^c}a_{ij}^{cd}(\eta_i(t)-\eta_{i,j}^c(t))) \\
  &\widehat{V}_i(t)=C_1\cb{X}_{i1}(t)
  \end{aligned}
\right.\\
&\text{and} \nonumber \\
&\left\{
\begin{aligned}
&\dot{\cb{X}}_{i2}(t)=A_2\cb{X}_{i2}(t)+B_2(\gamma\sum_{j \in \mathcal{N}_i^c}a_{ij}^{cd}(\widehat{V}_i(t)-\widehat{V}_{i,j}^c(t))) \\
&\eta_i(t)=C_2\cb{X}_{i2}(t)
\end{aligned}
\right.,
\end{alignat}
\end{subequations}}where $V_i(t)$ is the input signal to the DAC estimator, $\widehat{V}_i(t)\in \mathbb{R}$ is the estimated APV, and $\cb{X}_{i1}(t) \in \mathbb{R}^{n_1}, \cb{X}_{i2}(t) \in \mathbb{R}^{n_2}$ are the internal states of DAC estimator. Here $n_1$ and $n_2$ are positive integers, and {\color{black}$a_{ij}^{cd}>0$ is the DAC edge weight of $(i,j)\in\mathcal{E}_c$.} Moreover, $\eta_{i,j}^c(t), \widehat{V}_{i,j}^c(t)$ denote the required information from DGU $j$, and matrices $A_1, B_1, C_1, A_2, B_2, C_2$ and scalar $\gamma>0$ are DAC parameters invariant among all DGUs. The DAC estimator \eqref{eq:DAC estimator} achieves \textbf{Robust Average Consensus} (RAC) if $\widehat{V}_i(t)$ tracks the APV with zero steady-state error, i.e., 
{\begin{align}\label{eq:robust average consensus}
\widehat{V}_i(\infty)-\langle \cb{v}(\infty) \rangle =0,
\end{align}}regardless of the initial internal states $\cb{X}_{i1}(0), \cb{X}_{i2}(0),\forall i \in \mathcal{V}$. It is worth noting that RAC plays a vital role in supporting the plugging-in/out operations in DCmGs, as these operations will inevitably incur nonzero initial internal states for DAC estimators. Referring to Theorem 2 of \cite{bai2010robust}, we obtain the following result for \eqref{eq:DAC estimator}.

\begin{Lemma}\label{propos:design of h(s) and g(s)}
Under Assumption \ref{Ass:graph} and PCC voltages satisfying $V_i(s)=\frac{c_i^c(s)}{s}+\frac{c_i^r(s)}{s^2}, \forall i \in \mathcal{V}$, \footnote{$V_i(s)$ denotes the laplace transform of $V_i(t)$.} where $c_i^c(s)$ and $c_i^r(s)$ are polynomials that may differ among DGUs, all DAC estimators \eqref{eq:DAC estimator} in the DCmG can achieve RAC if
{\begin{subequations}\label{eq:h(s) and g(s)}
\begin{alignat}{2}
&h(s)=C_1(s{\rm I}^{n_1}-A_1)^{-1}B_1=\frac{2as+a^2}{(s+a)^2},\\
&g(s)=C_2(s{\rm I}^{n_2}-A_2)^{-1}B_2=\frac{s+a}{s^2},
\end{alignat}
\end{subequations}}where $a>0$ is an arbitrary scalar. Moreover, the following minimal realizations for $h(s)$ and $g(s)$ are adopted, i.e., 
{\begin{subequations}\label{eq:minimal realizations of LTIs}
\begin{alignat}{2}
&A_1=\left[ 
\begin{array}{cc}
-2a & -a^2 \\
1 & 0 
\end{array} 
\right],
B_1=\left[ 
\begin{array}{c}
1 \\
0 
\end{array} 
\right],
C_1=\left[ 
\begin{array}{cc}
2a & a^2
\end{array} 
\right],\\
&A_2=\left[ 
\begin{array}{cc}
0 & 0 \\
1 & 0 
\end{array} 
\right],\quad\quad\ 
B_2=\left[ 
\begin{array}{c}
1 \\
0 
\end{array} 
\right],\ 
C_2=\left[ 
\begin{array}{cc}
1 & a
\end{array} 
\right].
\end{alignat}
\end{subequations}}
\end{Lemma}
\begin{Proof}
{\color{black}The proof is given in Appendix \ref{appendix:proof of lemma 1}.}
\end{Proof}

\begin{Remark}
The statement in Lemma \ref{propos:design of h(s) and g(s)} also holds when $V_i(s)=\frac{c_i^c(s)}{q_i^c(s)s}+\frac{c_i^r(s)}{q_i^r(s)s^2}, \forall i \in \mathcal{V}$, where $q_i^c(s),q_i^r(s)$ are stable polynomials with all roots lying in the open left half-plane and contribute exponentially vanishing components to $V_i(s)$. According to Theorems \ref{theorem:attack impacts}-\ref{Theorem:Cooperative attack imapcts}, under ZTS attacks with constant $\bar{\cb{d}}_{ij}^a,(i,j)\in \tilde{\mathcal{E}}_c$, PCC voltages will eventually converge to stable values ($V_i(s)=\frac{c_i^c(s)}{q_i^c(s)s}$) or grow like ramp signals ($V_i(s)=\frac{c_i^r(s)}{q_i^r(s)s^2}$). Hence, under ZTS attacks satisfying Assumption \ref{Ass:consatnt faked unknown inputs}, all DAC estimators \eqref{eq:DAC estimator} with parameters set as \eqref{eq:minimal realizations of LTIs} can achieve RAC. Here we consider the minimal realizations for $h(s)$ and $g(s)$ as they require the minimum number of internal states in the DAC estimator \eqref{eq:DAC estimator}.
\end{Remark}

{\color{blue}To evaluate the integrity of the DAC related information communicated between DGUs $i$ and $j$, the UIO-based detectors are deployed. Specifically, DGU $i$ will utilize the received output information from DGU $j$ to estimate the internal states of the DAC estimator, and then compute residuals to detect possible attacks. Nevertheless, in terms of the estimator dynamics \eqref{eq:DAC estimator} with parameters set as \eqref{eq:minimal realizations of LTIs}, the UIO-based detectors are unable to detect any attack due to ${\rm rank}(B_1)={\rm rank}(C_1)$ and ${\rm rank}(B_2)={\rm rank}(C_2)$ \cite{gallo2020distributed}. That is, the number of decoupled unknown inputs is equal to the number of received independent outputs, and thus FDI attacks on the communicated outputs are indistinguishable from those caused by unknown inputs. To this end, to enable the attack detection ability of the UIO-based detectors, DGU $j$ will transmit the internal states $\cb{X}_{j1}(t)$ and $\cb{X}_{j2}(t)$ to DGU $i$, such that the number of received independent outputs can be increased. Moreover, the performance of the DAC estimator in tracking the APV is not affected as the required information $\widehat{V}_{i,j}^c(t)$ and $\eta_{i,j}^c(t)$ can be calculated from the received internal states $\cb{X}_{i,j1}^c(t)$ and $\cb{X}_{i,j2}^c(t)$, respectively.

\begin{Lemma}\label{lemma:UIOs for DAC estimators}
The integrity of the communicated data $\cb{X}_{i,j1}^c(t)$ and $\cb{X}_{i,j2}^c(t)$ is guaranteed via the following UIO-based detectors, i.e.,
{\begin{subequations}\label{eq:UIO for DAC estimators}
\begin{alignat}{2}
{\rm UIO}_{i,j}^{v}\left\{
\begin{aligned}
&\dot{\cb{z}}_{i,j}^{v}(t)=F_{j}^{v}\cb{z}_{i,j}^{v}(t)+\hat{{K}}_{j}^{v}\cb{X}_{i,j1}^c(t)\\
&\widehat{\cb{X}}_{i,j1}(t)=\cb{z}_{i,j}^{v}(t)+{H}_{j}^{v}\cb{X}_{i,j1}^c(t)
\end{aligned}
\right.,\\
{\rm UIO}_{i,j}^{\eta}\left\{
\begin{aligned}
&\dot{\cb{z}}_{i,j}^{\eta}(t)=F_{j}^{\eta}\cb{z}_{i,j}^{\eta}(t)+\hat{{K}}_{j}^{\eta}\cb{X}_{i,j2}^c(t)\\
&\widehat{\cb{X}}_{i,j2}(t)=\cb{z}_{i,j}^{\eta}(t)+{H}_{j}^{\eta}\cb{X}_{i,j2}^c(t)
\end{aligned}
\right.,
\end{alignat}
\end{subequations}}such that detection residuals
{ \begin{subequations}\label{eq:healthy component of residual for DAC}
\begin{alignat}{2}
&\cb{r}_{i,j1}^{v}(t)=\cb{X}_{i,j1}^c(t)-\widehat{\cb{X}}_{i,j1}(t)=e^{F_j^{v}t}\cb{\upepsilon}_{i,j1}^{v}(0),\\
&\cb{r}_{i,j2}^{\eta}(t)=\cb{X}_{i,j2}^c(t)-\widehat{\cb{X}}_{i,j2}(t)=e^{F_j^{\eta}t}\cb{\upepsilon}_{i,j2}^{\eta}(0),
\end{alignat}
\end{subequations}}both decay exponentially to zero in the absence of attacks. Here the UIO parameters $F_{j}^{v},\hat{{K}}_{j}^{v},{H}_{j}^{v}$ and $F_{j}^{\eta},\hat{{K}}_{j}^{\eta},{H}_{j}^{\eta}$ are set according to \eqref{eq:UIO parameters} to ensure that $F_j^{v}, F_j^{\eta}$ are both Hurwitz stable. $\cb{\upepsilon}_{i,j1}^{v}(0)$ and $\cb{\upepsilon}_{i,j2}^{\eta}(0)$ are the initial state estimation errors.
\end{Lemma}}
\begin{Proof}
{\color{black}The proof is given in Appendix \ref{appendix:proof for lemma 2}.}
\end{Proof}

Similar to \eqref{eq:normal condition 2}, the time-varying detection thresholds can be calculated such that 
{\begin{subequations}\label{eq:attack triggering condition of UIO for DAC}
\begin{alignat}{2}
&|\cb{r}_{i,j1}^{v}(t)| \le \bar{\cb{r}}_{i,j1}^{v}(t)=\kappa^{v} e^{-\mu^{v} t}\cb{\bar{\upepsilon}}_{i,j1}^{v}(0)\label{eq:attack triggering condition of UIO for DAC 1} \\
&|\cb{r}_{i,j2}^{\eta}(t)| \le \bar{\cb{r}}_{i,j2}^{\eta}(t)=\kappa^{\eta} e^{-\mu^{\eta} t}\cb{\bar{\upepsilon}}_{i,j2}^{\eta}(0) \label{aeq:attack triggering condition of UIO for DAC 2}
\end{alignat}
\end{subequations}}always hold in the absence of attacks. Once \eqref{eq:attack triggering condition of UIO for DAC 1} or \eqref{aeq:attack triggering condition of UIO for DAC 2} is violated, it is considered that the received $\cb{X}_{i,j1}^c(t)$ or $\cb{X}_{i,j2}^c(t)$ from DGU $j\in \mathcal{N}_i^c$ is corrupted.

{\color{blue}
Nevertheless, the attacker is still able to construct ZTS-like attacks to bypass the UIO-based detectors \eqref{eq:UIO for DAC estimators}, once she/he has full knowledge of the DAC parameters $A_1,B_1,C_1,A_2,B_2,C_2$, which are completely determined by the scalar $a$. According to the attack model, the attacker is able to obtain some system parameters including $a$ from insiders every few hours or days. Hence, based on the moving target defense (MTD) strategy, whose basic idea is to proactively perturb system parameters to make attacker's understanding of the system model outdated \cite{zhang2019analysis}, we assume that

\begin{Assumption}\label{Ass:MTD Strategy}
The DAC parameters $A_1,B_1,C_1,A_2,B_2, C_2$ can be hidden from the attacker.
\end{Assumption}

\begin{Remark}\label{remark:justifciation of the MTD strategy}
To hide the DAC parameters from the attacker, the perturbation strategy should be designed such that 1) the attacker cannot obtain the explicit perturbation command on $a$, which is denoted by $\Delta a$; 2) the attacker cannot infer $\Delta a$ from available information immediately. The first objective can be achieved by transmitting $\Delta a$ through encryption-based secure channels, while the second objective requires to perturb some extra parameters besides $a$ as the identification of transfer functions $h(s)$ and $g(s)$ (i.e., $a$) is possible through collecting the inputs and outs of the two linear dynamical systems involved in \eqref{eq:DAC estimator}. Hence, we choose to additionally perturb the DAC edge weights $a_{ij}^{cd},\forall (i,j)\in \mathcal{E}$ to hinder the identification of $h(s)$ and $g(s)$, as the inference of $a_{ij}^{cd}$ is usually time-consuming \cite{segarra2017network}. Therefore, if the control center can transmit the perturbation commands on $a$ and $a_{ij}^{cd}$ to all DGUs through secure channels every 5/10 minutes, then Assumption \ref{Ass:MTD Strategy} would be achieved. Moreover, it is noted that the perturbation will not impact the robust average consensus once $a>0$ and $a_{ij}^{cd}>0$ are guaranteed.
\end{Remark}}

\subsection{Attack Detection Phase}
In this subsection, we introduce the detection indicator and the corresponding detection threshold, under which the detectability for ZTS attacks is investigated. Through comparing the nominal reference voltage $V_{ref}$ with the estimated APV $\widehat{V}_i(t)$, we obtain the estimated average PCC voltage deviation (APVD) as
{\begin{align}\label{eq:estimated average PCC voltage deviation}
\widehat{V}_i^{err}(t)=V_{ref}-\widehat{V}_i(t).
\end{align}}

{\color{blue}Although daily operations (e.g., load switches and plugging-in/out of DGUs) in DCmGs never cause steady-state APVD, i.e., $\langle \cb{v}(\infty) \rangle = V_{ref}$, non-trivial instantaneous APVD will emerge as it takes some time for $\langle \cb{v}(t) \rangle$ to converge as Fig. \ref{fig:simulation resultss for the set of detection thrsholds} shows. Thus, both daily operations and ZTS attacks will lead to non-trivial $\widehat{V}_i^{err}(t)$, and it may be difficult to distinguish attacks from daily operations based on only historical and current non-trivial $\widehat{V}_i^{err}(t)$. Fortunately, we observe that the non-trivial $|\widehat{V}_i^{err}(t)|$ caused by daily operations shares a common characteristic, i.e., $|\widehat{V}_i^{err}(t)|$ reaches its peak value at almost the time when the daily operations occur and then it will quickly decay to zero. Differently, under Assumption \ref{Ass:consatnt faked unknown inputs}, the ZTS attack \eqref{eq:dynamics of ZTS} will cause either constant or ramp-growing APVD. Thus, it is natural to derive the following detection indicator by utilizing the technology of sliding time window (STW).}
\begin{Definition}[STW-based Detection Indicator]
Given the time window with fixed length $T$, the detection indicator $\mathcal{d}_i(t)$ is computed as the integral of the time window $(t-T,t)$ sliding over $|\widehat{V}_i^{err}(t)|$, i.e.,
{\begin{align}\label{Calculation d_i}
\mathcal{d}_i(t)=\left\{
\begin{aligned}
&0, t_s+T>t\ge t_s \\
&\int_{t-T}^t|\widehat{V}_i^{err}(\tau)|{\rm d}\tau,t\ge t_s+T
\end{aligned}
\right.,
\end{align}}where $t_s >0$ is the activation time for the generation of $\mathcal{d}_i(t)$.
\end{Definition}

Next, we investigate the setting of the detection threshold under which certain daily operations can be tolerated. Let $\mathcal{O}(t)=\{o_1(t),\cdots,o_{|\mathcal{O}|}(t)\}$ be the set of daily operations, where $o_k(t) \in \mathcal{O}(t)$ represents the event of a daily operation occurring at time $t$. To tolerate any single daily operation contained in $\mathcal{O}(t),\forall t \ge t_s $, the constant detection threshold is set as
{\begin{align}\label{eq:detection threshold}
\bar{\mathcal{d}}_i=\max_{\substack{t\ge t_s+T \\ o_k(t_s) \in \mathcal{O}(t_s)}} \int_{t-T}^t|\widehat{V}_{i|o_k(t_s)}^{err}(\tau)|{\rm d}\tau,t\ge t_s+T,
\end{align}}where $\widehat{V}_{i|o_k(t_s)}^{err}(t)$ denotes the estimated APVD under the daily operation $o_k(t_s)$, and could be obtained from the historical real world data or the simulated data. To preserve the detectability for ZTS attacks, it is suggested to tolerate the most frequent daily operations in DCmGs. Thus, under any daily operation $o_k(t) \in \mathcal{O}(t), \forall t \ge t_s$, we get
{\begin{align}\label{eq:normal condition}
\mathcal{d}_i(t)\le\bar{\mathcal{d}}_i,\forall t \ge t_s.
\end{align}}

Once \eqref{eq:normal condition} is violated, it is considered that there exist ZTS attacks and the impact mitigation phase is activated.
\begin{Theorem}\label{propos: detectability for ZTS and ZTS attacks}
Under Assumptions \ref{Ass:Voltage}-\ref{Ass:MTD Strategy} and the DAC estimators \eqref{eq:DAC estimator} satisfying \eqref{eq:minimal realizations of LTIs}, ZTS attacks \eqref{eq:dynamics of ZTS} can be detected by the STW-based detection indicator $\mathcal{d}_i(t)$ if 
{\begin{align}\label{eq:detectability condition}
\langle \cb{\psi}_a(\infty) \rangle >\frac{1}{T}\bar{\mathcal{d}}_i,
\end{align}}where $\langle \cb{\psi}_a(t) \rangle$ denotes the APVD caused by ZTS attacks.
\end{Theorem}
\begin{Proof}
{\color{black}The proof is given in Appendix \ref{appendix:proof of Theorem 4}.}
\end{Proof}

\begin{Remark}
Here the detection threshold $\bar{\mathcal{d}}_i$ does not increase linearly with the time window length $T$, as the estimated APVD will eventually converge to zero, i.e., $\widehat{V}_{i|o_k(t_s)}^{err}(\infty)=0$. Thus, according to \eqref{eq:detectability condition}, the detectability for ZTS attacks with constant $\bar{\cb{d}}_{ij}^a$ would be enhanced with a larger $T$. That is, a larger $T$ could decrease the impact of daily operations on the detectability for ZTS attacks. Meanwhile, we should also be aware of that the STW technology will result in certain amount of initial detection delay and some computation burden for each DGU, and thus $T$ cannot be set arbitrarily large.
\end{Remark}

\vspace{-10pt}
\subsection{Impact Mitigation Phase}
In this subsection, we introduce the impact mitigation phase that is activated once \eqref{eq:normal condition} is violated. In particular, let $t=T_i^{alm}$ be the time when $\mathcal{d}_i(T_i^{alm})>\bar{\mathcal{d}}_i$, after which the estimated APVD $\widehat{V}_i^{err}(t)$ is fed into the proportional-integral (PI) based compensator, i.e.,
{\begin{align}\label{eq:Compensator}
C_i(t)=k_{cp}\widehat{V}_i^{err}(t)+k_{ci}\int_{T_i^{alm}}^t \widehat{V}_i^{err}(\tau){\rm d}\tau,
\end{align}}where $k_{cp}>0$ and $k_{ci}>0$ are PI compensation gains, and the compensation value $C_i(t)$ will be added to the secondary control input $\psi_i(t)$. With some abuse of the notation, the compensated secondary control input $\psi_i(t)$ is decomposed as
{\begin{align}\label{eq:impact counteraction}
\psi_i(t)=\tilde{\psi}_i(t)+\psi_i^a(t)+C_i(t),
\end{align}}where $\tilde{\psi}_i(t)$ is the healthy component, and $\psi_i^a(t)$ is the malicious component associated with attacks.

\begin{Theorem}\label{propos:effectiveness of the counteraction}
Under Assumptions \ref{Ass:Voltage}-\ref{Ass:MTD Strategy} and the DAC estimators \eqref{eq:DAC estimator} satisfying \eqref{eq:minimal realizations of LTIs}, if \eqref{eq:normal condition} is violated, then the activated impact mitigation strategy \eqref{eq:impact counteraction} can eventually achieve
{\begin{align}\label{eq:average PCC voltage after compensation}
\langle \cb{v}(\infty) \rangle=V_{ref}-\frac{1}{k_{ci}} \sum_{(i,j) \in \tilde{\mathcal{E}}_c}\frac{k_Ia_{ij}^c}{NI_{tj}^s}\cb{k}^{\rm T}A_{kj}^{-1}\bar{E}_j\bar{\cb{d}}_{ij}^a.
\end{align}}
\end{Theorem}

\begin{Proof}
{\color{black}The proof is given in Appendix \ref{appendix:proof of theorem 5}.}
\end{Proof}

\begin{Remark}\label{Remark:impact mitigation}
For the cooperative ZTS attacks satisfying \eqref{eq:cooperative attacks conditions}, the impact mitigation strategy \eqref{eq:impact counteraction} can eliminate the constant APVD caused by them. While regarding to the non-cooperative attacks where \eqref{eq:cooperative attacks conditions} is not satisfied and the ramp-growing APVD is induced, the impact mitigation strategy \eqref{eq:impact counteraction} can stabilize all PCC voltages with constant APVD, which is determined by \eqref{eq:average PCC voltage after compensation}. {\color{black} Moreover, we note that the impact mitigation strategy \eqref{eq:impact counteraction} will not destroy the voltage balancing when any daily operations falsely trigger the attack alarm \eqref{eq:normal condition} (i.e., the false alarm) once the PI compensation gains are well tuned, which is validated in the supplementary material.}
\end{Remark}


Once the detection indicator $\mathcal{d}_i(t)$ is smaller than a predefined threshold $T\delta>0$ at $t=T_i^{com}$, i.e.,
{\begin{align}\label{eq:compensation completed condition}
|\mathcal{d}_i(T_i^{com})|<T\delta,
\end{align}}it is considered that the voltage balancing has almost been recovered. Then, the impact mitigation strategy \eqref{eq:impact counteraction} is disabled, and the corresponding compensation value $C_i(T_i^{com})$ is added to the secondary control input $\psi_i(t)$ as a constant.

Although we only analyze the effectiveness of Algorithm \ref{Algorithm: the whoel countermeasure} under ZTS attacks with constant $\bar{\cb{d}}_{ij}^a$, it should be emphasized that Algorithm \ref{Algorithm: the whoel countermeasure} is also effective when $\bar{\cb{d}}_{ij}^a(t)$ is time-varying. 
In simulations, we show the effectiveness of Algorithm \ref{Algorithm: the whoel countermeasure} under the ZTS attack with sine signal $\bar{\cb{d}}_{ij}^a(t)$ in Fig. \ref{fig:systemstates under attack set IV}.

\section{Simulation Studies}\label{section:simulation}

In this section, we conduct extensive simulation studies on the DCmG composed of 8 DGUs established in Matlab Simulink/PLECS to validate the theoretical results. The corresponding electrical parameters are provided in Appendix \ref{appendix:electrical_parameyters}. The nominal reference voltage is set as $V_{ref}=48{\rm V}$, and the bounds of noises are $\bar{\cb{\rho}}_i=\bar{\cb{\omega}}_i=[0.001, 0.003, 0]^{\rm T}, \forall i \in \mathcal{V}=\{1,\cdots,8\}$. The weight parameter involved in \eqref{eq:secondary control input} is $k_I=5$. The DAC parameters are set according to \eqref{eq:minimal realizations of LTIs} with $a=100$, and the length of STW is $T=0.65$s. Moreover, the PI compensation gains in \eqref{eq:Compensator} are $k_{cp}=1,k_{ci}=20$, and the threshold judging the achievement of voltage balancing is set as $\delta=0.005$V.

\subsection{The Setting of $\bar{\mathcal{d}}_i$}
In this subsection, we investigate the setting of detection thresholds $\bar{\mathcal{d}}_i, \forall i \in \mathcal{V}$ that can tolerate any daily operation contained in the set $\mathcal{O}(t)=\{o_1(t),o_2(t),o_3(t)\}$, whose elements are elaborated in TABLE \ref{Table:Daily operation set}. Before implementing $\mathcal{O}(t)$, a series of initialization operations are conducted as indicated in Fig. \ref{fig:Evolutions_deetction threshodls}, which are introduced as follows: at $t=0{\rm s}$, all primary controllers are activated; at $t=2 {\rm s}$, all DGUs except DGU 7 are connected through power lines; at $t=4 {\rm s}$, the communication network is established, and UIO-based detectors \eqref{eq:UIO}, DAC estimators \eqref{eq:DAC estimator}, and the generation of $\mathcal{d}_i(t)$ are activated. Then, the daily operations are introduced: at $t=8 {\rm s}$, DGU $8$ is plugged out from the DCmG; at $t=12 {\rm s}$, all load currents are decreased by $30\%$ of their rated values; at $t=16{\rm s}$, DGUs 7 is plugged into the DCmG. 


\begin{table}[h]
\centering 
\caption{Elaboration of the daily operation set $\mathcal{O}(t)$} \label{Table:Daily operation set}
\scalebox{1.2}{
\begin{tabular}{|p{0.8cm}<{\centering}|p{6.5cm}<{\centering}|}
\hline
$o_1(t)$ & plugging out of DGU $8$ \\ \hline
$o_2(t)$ & decrease of load currents by $30\%$ of their rated values  \\ \hline
$o_3(t)$ & plugging in of DGU $7$ \\ 
\hline
\end{tabular}}
\end{table}

\begin{figure}[thpb]
  \centering
  \includegraphics[scale=0.25]{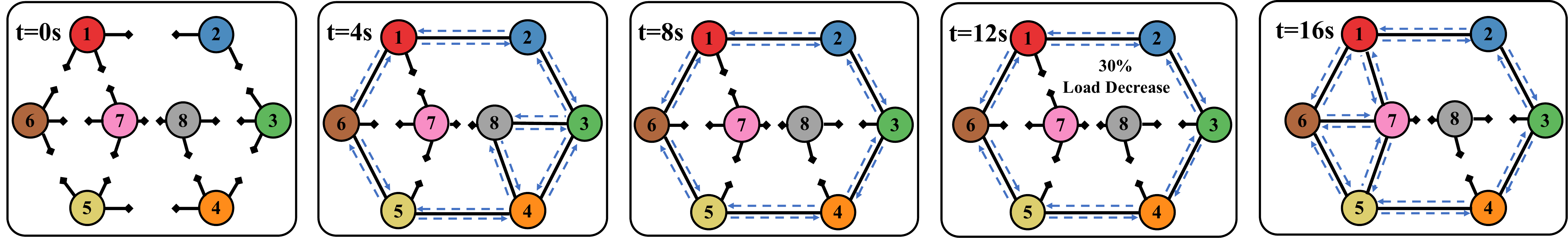} 
  \caption{This figure shows the evolution of the DCmG corresponding to the daily operation set $\mathcal{O}(t)$. Here the solid black lines are power lines and the dotted blue lines signify communication links.} \label{fig:Evolutions_deetction threshodls}
\end{figure}

As shown in Fig. \ref{fig:simulation resultss for the set of detection thrsholds}, each daily operation will cause non-trivial disturbance on the estimated APVD, and some fluctuation emerges on the detection indicator $\mathcal{d}_i(t)$ accordingly. Moreover, it is observed that any daily operation in $\mathcal{O}(t)$ can be tolerated by the detection thresholds ${\color{black}\bar{\mathcal{d}}_i}=0.0325,\forall i \in \mathcal{V}$. Hence, according to Theorem \ref{propos: detectability for ZTS and ZTS attacks}, under Assumption \ref{Ass:consatnt faked unknown inputs}, ZTS attacks causing the steady-state APVD more than $\frac{\bar{\mathcal{d}}_i}{T}=0.05$V will be detected.

\begin{figure}[thpb]
  \centering
  \includegraphics[scale=0.3]{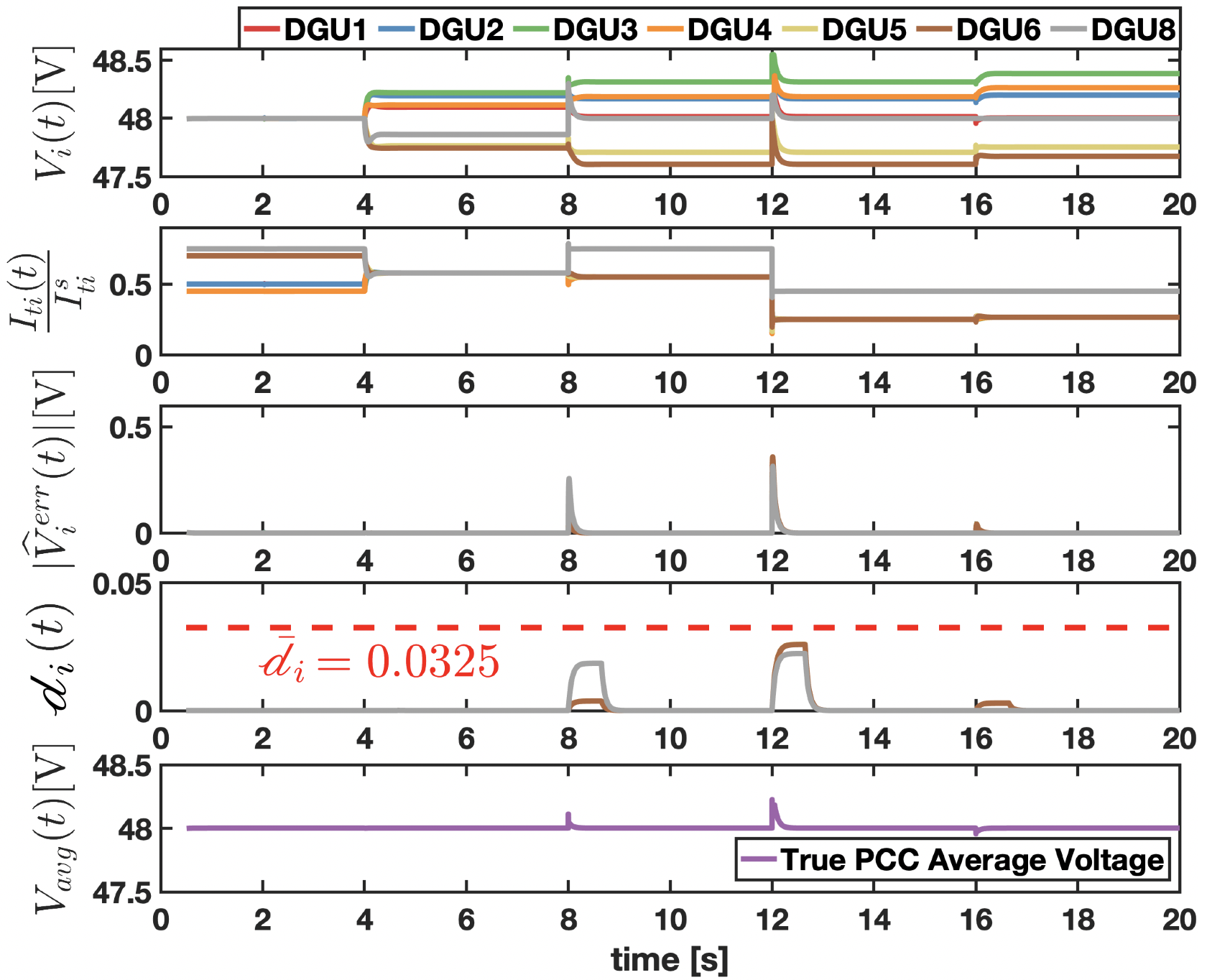}
  \caption{This figure shows PCC voltages $V_i(t)$, output currents in per-unit $\frac{I_{ti}(t)}{I_{ti}^s}$, estimated APVDs $\widehat{V}_i^{err}(t)$, detection indicators $\mathcal{d}_i(t),\forall i \in \mathcal{V}$ and the true average PCC voltage $V_{avg}(t)$.}\label{fig:simulation resultss for the set of detection thrsholds}
\end{figure}

\subsection{ZTS Attacks with Constant $\bar{\cb{d}}_{ij}^a$}
In this subsection, we validate the effectiveness of Algorithm \ref{Algorithm: the whoel countermeasure} against ZTS attacks with constant $\bar{\cb{d}}_{ij}^a$. In particular, we consider two cases where single ZTS attack and cooperative ZTS attacks are launched.

\subsubsection{Attack Set {\rm I}}
Attack set I is composed of one ZTS attack targeting at communication link $(8,3)$, and its attack vector is generated by \eqref{eq:dynamics of ZTS} with $\bar{\cb{d}}_{83}^{a1}=[2,0]^{\rm T}$. Attack set I is activated at $T_{a1}=6$s. As shown in (b) of Fig. \ref{fig:residuals under attack set II}, the detection residuals under attack set I are still bounded by the detection thresholds, indicating that attack set I is unforeseeable for UIO$_{8,3}$.

\begin{figure}[htbp]
\centering
\includegraphics[scale=0.25]{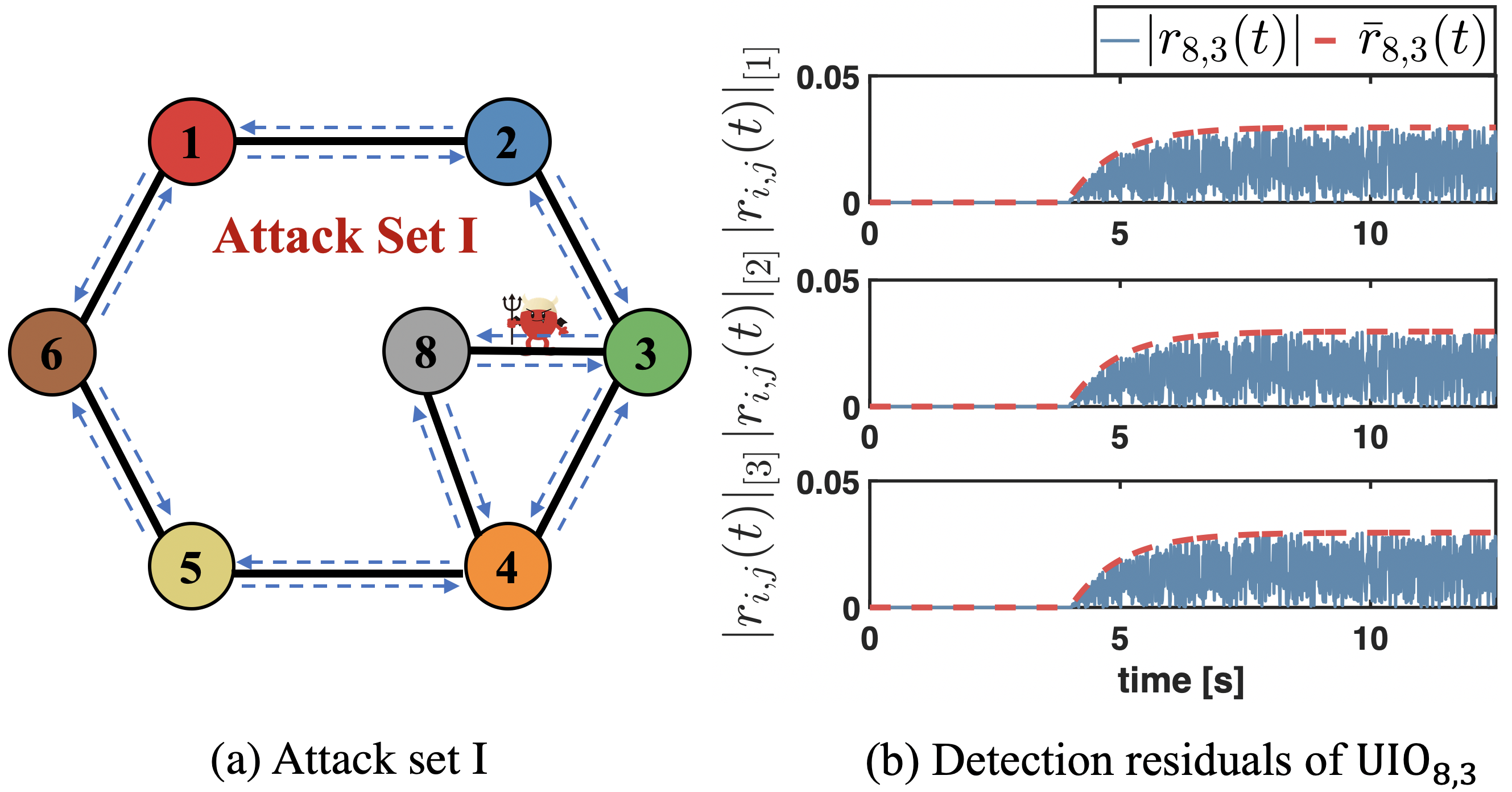}
\caption{This figure depicts the location of attack set I and the detection residuals of UIO$_{8,3}$ under attack set I.}\label{fig:residuals under attack set II}
\end{figure}
{\color{blue}According to Fig. \ref{fig:systemstates under attack set II}, attack set I incurs ramp-growing APVD and the current sharing is damaged. While the activated countermeasure can make the APV finally converge and the steady-state APVD is $0.017$V, which can significantly mitigate the attack impact and timely avoid the occurrence of a blackout incident in the DCmG.}

\begin{figure}[htbp]
\includegraphics[scale=0.3]{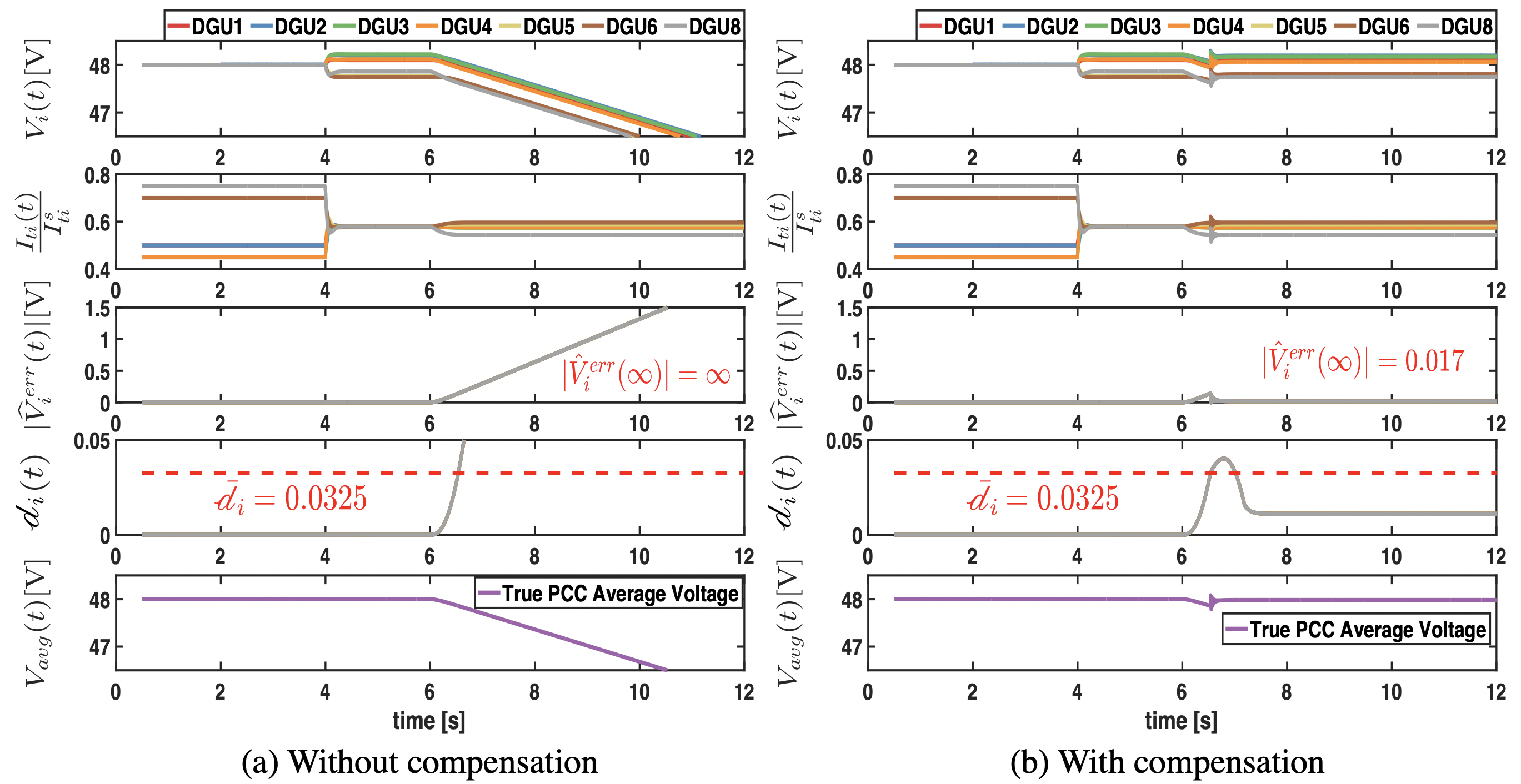}
\centering
\caption{This figure shows PCC voltages, output currents and the countermeasure related variables under attack set I without compensation ($k_{cp}=k_{ci}=0$) and with compensation ($k_{cp}=1,k_{ci}=20$).} \label{fig:systemstates under attack set II}
\end{figure}

\subsubsection{Attack Set {\rm II}}
Attack set II is composed of two cooperative ZTS attacks targeting at communication links $(2,1)$ and $(3,2)$, and corresponding attack vectors are generated by \eqref{eq:dynamics of ZTS} with parameters $\bar{\cb{d}}_{21}^{a2}=[2,0]^{\rm T}$ and $\bar{\cb{d}}_{32}^{a2}=[-2.8,0]^{\rm T}$, respectively. Attack set II is activated at $T_{a2}=6$s. Similarly, as shown in (b) of Fig. \ref{fig:residuals under attack set III}, attack set II can bypass the detection of UIO$_{2,1}$ and UIO$_{3,2}$, as the corresponding detection residuals are almost not impacted.

\begin{figure}[htbp]
\centering
\includegraphics[scale=0.25]{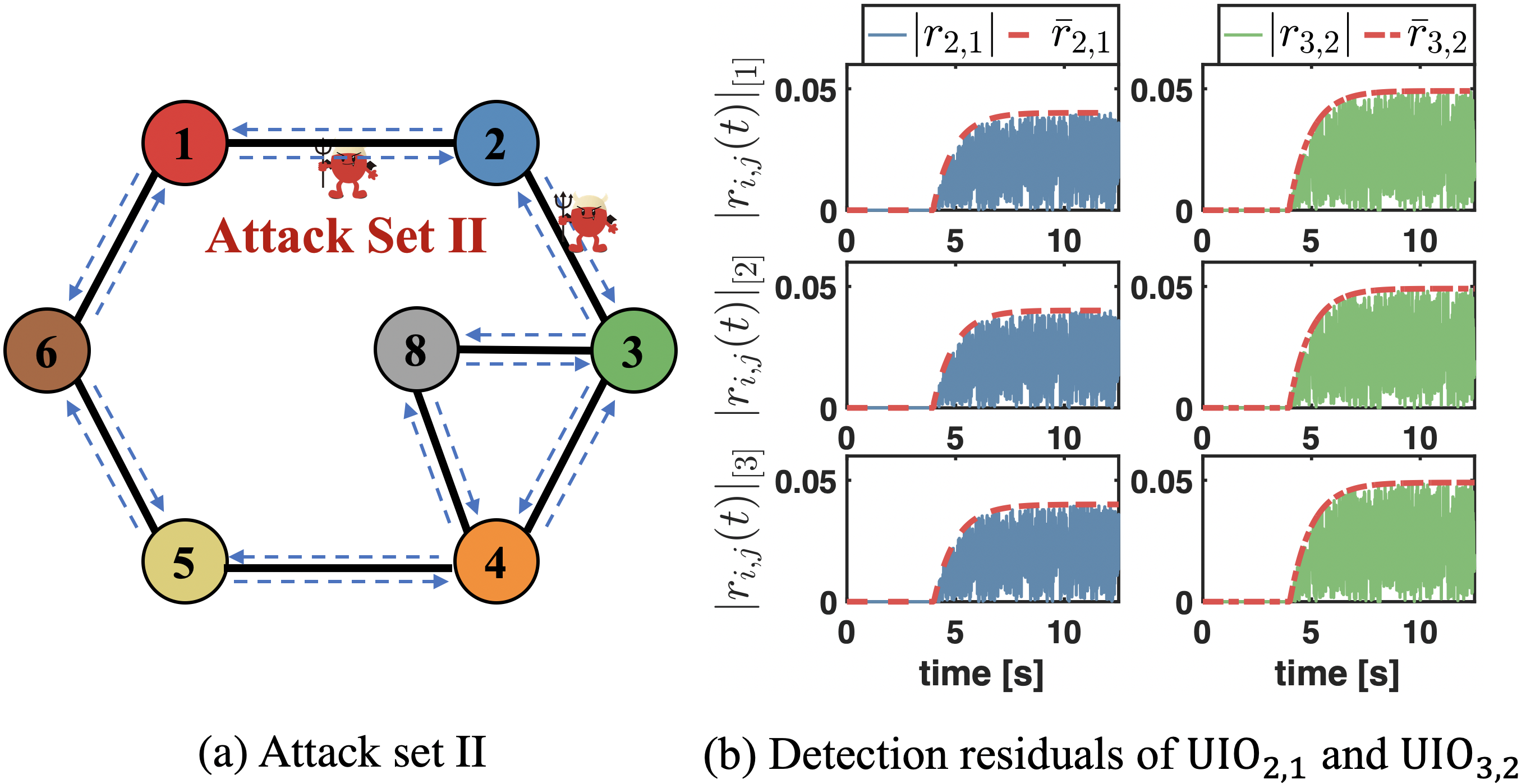}
\caption{This figure depicts the location of attack set II and the detection residuals of UIO$_{2,1}$ and UIO$_{3,2}$ under attack set II.}\label{fig:residuals under attack set III}
\end{figure}

{\color{blue}According to Fig. \ref{fig:systemstates under attack set III}, attack set II causes constant APVD and destroys the current sharing in DCmGs, which validates the correctness of Theorem \ref{Theorem:Cooperative attack imapcts}. The activated countermeasure can effectively eliminate the malicious APVD and pull the PCC voltages of all DGUs to the nominal reference point, which validates the statement in Remark \ref{Remark:impact mitigation}.}

\begin{figure}[htbp]
\includegraphics[scale=0.3]{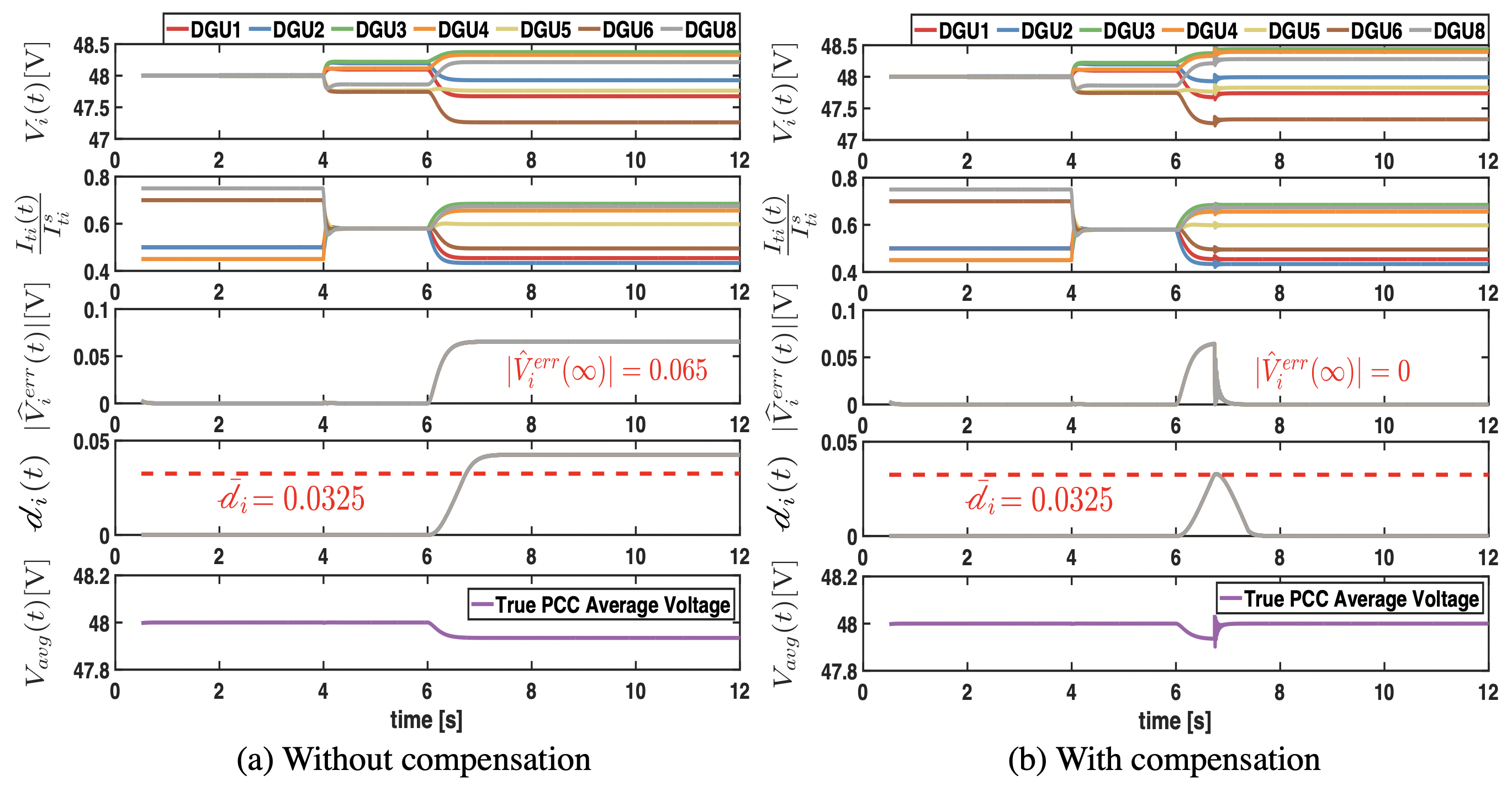}
\centering
\caption{This figure shows PCC voltages, output currents and countermeasure related variables under attack set II without compensation ($k_{cp}=k_{ci}=0$) and with compensation ($k_{cp}=1,k_{ci}=20$).} \label{fig:systemstates under attack set III}\vspace{-15pt}
\end{figure}

\subsection{ZTS Attack with Time-varying $\bar{\cb{d}}_{ij}^a(t)$}
In this subsection, the effectiveness of Algorithm \ref{Algorithm: the whoel countermeasure} against the ZTS attack with time-varying $\bar{\cb{d}}_{ij}^a(t)$ is shown. Attack set III is composed of one ZTS attack targeting at communication link $(8,3)$, and the attack vector is generated by \eqref{eq:dynamics of ZTS} with $\bar{\cb{d}}_{83}^{a3}(t)=[{\rm sin}(4t),0]^{\rm T}$. Attack set III is activated at $T_{a3}=6$s. According to Fig. \ref{fig:systemstates under attack set IV}, it is validated that the countermeasure can substantially decrease the APVD caused by attack set III, under which the APVD after compensation can be almost neglectable. 

\begin{figure}[htbp]
\includegraphics[scale=0.3]{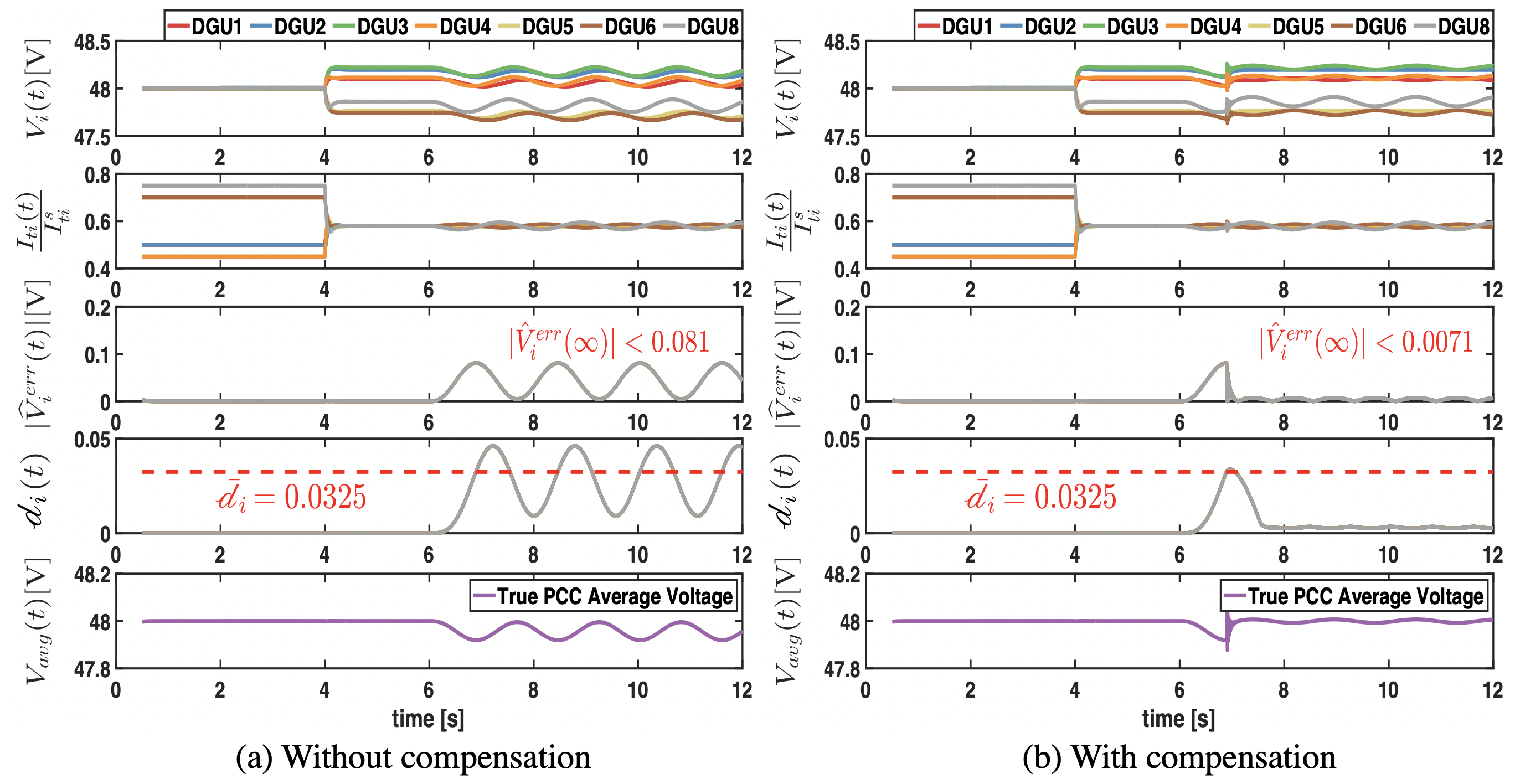}
\centering
\caption{This figure shows PCC voltages, output currents and countermeasure related variables under under attack set III without compensation ($k_{cp}=k_{ci}=0$) and with compensation ($k_{cp}=1,k_{ci}=20$).} \label{fig:systemstates under attack set IV}
\end{figure}

For clarity, we present TABLE \ref{table:2} to sum up the APVDs without compensation and with compensation under the three attack sets aforementioned.

\renewcommand\arraystretch{1.5}
\begin{table}[h]
\centering 
\caption{APVDs under the three attack sets}\label{table:2}
\scalebox{1.2}{
\begin{tabular}{|p{1.8cm}<{\centering}|p{2.8cm}<{\centering}|p{2.8cm}<{\centering}|}
\hline
 & Without compensation & With compensation \\ \hline
Attack set I & $|\widehat{V}_i^{err}(\infty)|=\infty$V & $|\widehat{V}_i^{err}(\infty)|=0.017$V \\ \hline
Attack set II & $|\widehat{V}_i^{err}(\infty)|=0.065$V & $|\widehat{V}_i^{err}(\infty)|=0$V \\ \hline
Attack set III & $|\widehat{V}_i^{err}(\infty)|\le0.081$V & $|\widehat{V}_i^{err}(\infty)|\le0.0071$V \\
\hline
\end{tabular}}
\end{table}

\section{Conclusion}\label{section:conclusion and future works}
{\color{blue}In this paper, we revealed that the potential vulnerability of the UIO-based detector originates from the lacked knowledge of true unknown inputs. ZTS attacks can be constructed by secretly faking the unknown inputs.} Moreover, it is proved that single ZTS attack can destabilize the DCmG, and cooperative ZTS attacks can cause accurate and specific impact, which could be manipulated by the attacker. Through estimating the APV, we design a distributed countermeasure against ZTS attacks, which could decrease the APVD or even recover the voltage balancing in DCmGs. {\color{black}In the future work, we will design a resilient control framework in DCmGs where voltage balancing and current sharing can be both attained in the presence of attacks, and rigorously investigate the PI compensation gains' stability region under which the extreme daily operations will not diverge the PCC voltages.}

\footnotesize{
\bibliographystyle{IEEEtran}
\bibliography{V4-Revisedpaper20210728-Singlecolumns}}

\normalsize
\appendix
\subsection{Proof of Theorem \ref{theorem:attack impacts}}\label{appendix:theorem 2}
After simplifying the primary control loops as unit gains \cite{tucci2018stable}, we have
\begin{align}\label{eq:simplified primary contorl loop}
\cb{v}(t)=\cb{v}_{r}+\cb{\psi}(t),
\end{align}where $\cb{v}_{r}=V_{ref}\mathbb{1}^N$ is the constant nominal reference PCC voltage vector. Moreover, integrating \eqref{eq:secondary control input} with \eqref{eq:attack model}, the dynamics of the secondary control input under attacks are
\begin{align}\label{eq:intergrated Secondary control input}
\dot{\cb{\psi}}(t)=-\tilde{L}D\cb{i}_t(t)+\frac{k_Ia_{ij}^c}{I_{tj}^s}\cb{k}^{\rm T}\cb{\phi}_{i,j}(t)\cb{l}_i,
\end{align}where $\tilde{L}=k_IL, D=$ diag$\{\frac{1}{I_{t1}^s},\cdots,\frac{1}{I_{tN}^s}\}$, and $\cb{l}_i \in \mathbb{R}^N$ is obtained from $\mathbb{0}^N$ with its $i$-th element replaced by 1. Here $\cb{i}_t(t)=[I_{t1}(t),\cdots,I_{tN}(t)]^{\rm{T}}$ is the output current vector under attacks and, according to the Kirchhoff current law, we obtain
\begin{align}\label{eq:24}
\cb{i}_t(t)=M\cb{v}(t)+\cb{i}_l,
\end{align}where $\cb{i}_l=[I_{L1},\cdots,I_{LN}]^{\rm{T}}$ is the constant load current vector. The overall dynamics of the hierarchical control framework can be obtained after integrating equations \eqref{eq:simplified primary contorl loop}-\eqref{eq:24}, i.e.,
\begin{align}\label{eq:dynamcis of the seconadry control input under attacks}
\dot{\cb{\psi}}(t)=-Q\cb{\psi}(t)-\tilde{L}D\cb{i}_l-Q\cb{v}_{r}+\frac{k_Ia_{ij}^c}{I_{tj}^s}\cb{k}^{\rm T}\cb{\phi}_{i,j}(t)\cb{l}_i,
\end{align}where $Q=\tilde{L}DM$ integrates the Laplacian matrices of graphs $\mathcal{G}_c$ and $\mathcal{G}_{el}$. According to Proposition 3 in \cite{tucci2018stable}, $Q$ satisfies
\begin{enumerate}
  \item[a)] {\rm ker}$(Q)$=$\mathbb{H}_{\bot}^1$, {\rm range}$(Q)$= $\mathbb{H}^1$;
  \item[b)] $Q$ is diagonalizable and has non-negative eigenvalues, and its algebraic multiplicity of zero eigenvalue is one.
\end{enumerate}
Hence, eigenvalue eigenvector pairs of $Q$ can be denoted by $\cb{p}_i=(\lambda_i, \cb{q}_i), \forall i \in \mathcal{V}$, where $\lambda_1=0$, $0<\lambda_2\le\cdots\le\lambda_N$, $\cb{q}_i \in \mathbb{R}^N$, $\cb{q}_1 \in \mathbb{H}_\bot^1$, and $\{\cb{q}_2,\cdots,\cb{q}_N\}$ constitutes a basis of $\mathbb{H}^1$. Given the linear differential equation \eqref{eq:dynamcis of the seconadry control input under attacks}, the healthy component of the secondary control input vector can be decomposed and calculated as
{\begin{align}{\label{eq:21}}
\cb{\tilde{\psi}}(t)=e^{-Qt}\tilde{\cb{\psi}}(0)+\sum_{i=2}^N\frac{\alpha_i}{\lambda_i}(1-e^{-\lambda_it})\cb{q}_i,
\end{align}}where $\alpha_i\in\mathbb{R},\forall i \in \mathcal{V}$ are chosen such that $\sum_{i=2}^N\alpha_i\cb{q}_i=-\tilde{L}D\cb{i}_l-Q\cb{v}_{r} \in \mathbb{H}^1$. 

Under Assumption \ref{Ass:consatnt faked unknown inputs}, the ZTS attack vector can be expressed as
\begin{align}\label{Attack vector of the ZTS attack}
\cb{\phi}_{i,j}(t)=e^{A_{kj}(t-T_a)}\check{\cb{\phi}}_{i,j}(T_a)-A_{kj}^{-1}\bar{E}_j\bar{\cb{d}}_{ij}^a,
\end{align}where $\check{\cb{\phi}}_{i,j}(T_a)=A_{kj}^{-1}\bar{E}_j\bar{\cb{d}}_{ij}^a$. The malicious component of the secondary control input vector is decomposed as $\cb{\psi}_{a}(t)=\cb{\psi}_{a1}(t)+\cb{\psi}_{a2}(t)$, where $\cb{\psi}_{a1}(t)$ and $\cb{\psi}_{a2}(t)$ represent the components associated with $e^{A_{kj}(t-T_a)}\check{\cb{\phi}}_{i,j}(T_a)$ and $-A_{kj}^{-1}\bar{E}_j\bar{\cb{d}}_{ij}^a$, respectively. As $A_{kj}^{-1}\bar{E}_j\bar{\cb{d}}_{ij}^a$ is a constant vector, $\cb{\psi}_{a2}(t)$ can be directly calculated as
\begin{align}\label{eq:constant part impact on secondary input}
\cb{\psi}_{a2}(t)  = \alpha_{1l}(t-T_a)\cb{q}_1+\sum_{i=2}^N\frac{\alpha_{il}}{\lambda_i}(1-e^{-\lambda_i(t-T_a)})\cb{q}_i,
\end{align}where $\alpha_{il} \in \mathbb{R},\forall i \in \mathcal{V}$ are chosen such that $\sum_{i=1}^N\alpha_{il}\cb{q}_i=-\frac{k_Ia_{ij}^c}{I_{tj}^s}\cb{k}^{\rm T}A_{kj}^{-1}\bar{E}_j\bar{\cb{d}}_{ij}^a\cb{l}_i$. For brevity, we only show the case where $\cb{A_{kj}}$ is diagonalizable\footnote{When $A_{kj}$ is not diagonalizable, the stability and convergence properties can be analyzed in a similar way.}, with which we obtain $e^{A_{kj}(t-T_a)}\check{\cb{\phi}}_{i,j}(T_a)=\sum_{m=1}^3\eta_me^{\beta_m(t-T_a)}\cb{a}_m$, where $(\beta_m,\cb{a}_m), \forall m \in \underline{\mathcal{V}}=\{1,2,3\}$ are eigenvalue eigenvector pairs of $A_{kj}$, and $\eta_m,\forall m \in \underline{\mathcal{V}}$ are scalars such that $\sum_{m=1}^3\eta_m\cb{a_m}=\check{\cb{\phi}}_{i,j}(T_a)$. Then, $\cb{\psi}_{a1}(t)$ is calculated as
{\begin{align}\label{eq:first}
\cb{\psi}_{a1}(t)=\sum_{m=1}^3\frac{k_Ia_{ij}^c}{I_{tj}^s}\eta_ma_{m2}e^{-\beta_mT_a}e^{-Qt}\int_{T_a}^t e^{\beta_m\tau}e^{Q\tau}\cb{l}_id\tau,
\end{align}}where $a_{m2} = \cb{k}^{\rm T}\cb{a}_m \in \mathbb{R}$ denotes the second entry of vector $\cb{a}_m$. As the set $\{\cb{q}_1,\cb{q}_2,\cdots,\cb{q}_N\}$ constitutes a basis of $\mathbb{R}^N$, $\cb{l}_i$ can be expressed as $\cb{l}_i=\sum_{r=1}^{N}\delta_r\cb{q}_r$, where $\delta_1\cb{q}_1=\langle\cb{l}_i\rangle\mathbb{1}^N=\frac{1}{N}\mathbb{1}^N$. Thus, we obtain $e^{Qt}\cb{l}_i=\sum_{r=1}^N\delta_re^{\lambda_rt}\cb{q}_r$. Similarly, we only show the case where $\beta_m+\lambda_r \ne 0, \forall m \in \underline{\mathcal{V}}, r \in \mathcal{V}$,\footnote{When $\beta_m=-\lambda_r$, the integral part in \eqref{eq:integral} can be calculated similarly.} then the integral component in \eqref{eq:first} is computed as
{\begin{align}\label{eq:integral}
\int_{T_a}^t e^{\beta_m\tau}e^{Q\tau}\cb{l}_id\tau=\sum_{r=1}^N\frac{\delta_r}{\beta_m+\lambda_r}(e^{(\beta_m+\lambda_r)t}-e^{(\beta_m+\lambda_r)T_a})\cb{q}_r.
\end{align}}Substituting \eqref{eq:integral} into \eqref{eq:first}, we have
{\begin{align}\label{eq:impacts of variation part on secondary control input}
\cb{\psi}_{a1}(t)=\sum_{m=1}^3\sum_{r=1}^N\frac{k_Ia_{ij}^c\eta_ma_{m2}\delta_r}{I_{tj}^s(\beta_m+\lambda_r)}(e^{\beta_m(t-T_a)}-e^{-\lambda_r(t-T_a)})\cb{q}_r.
\end{align}}

With $\alpha_{1l}\cb{q}_1=-\frac{k_Ia_{ij}^c}{NI_{tj}^s} \cb{k}^{\rm T}A_{kj}^{-1}\bar{E}_j\bar{\cb{d}}_{ij}^a\mathbb{1}^N$, the average of elements in $\cb{\psi}_{a2}(\infty)$ is calculated as
\begin{align}\label{eq:Inifity of constant part}
\langle \cb{\psi}_{a2}(\infty) \rangle = \langle \alpha_{1l}\cb{q}_1 \rangle(t-T_a)
=-\frac{k_Ia_{ij}^c}{NI_{tj}^s}\cb{k}^{\rm T}A_{kj}^{-1}\bar{E}_j\bar{\cb{d}}_{ij}^a(t-T_a).
\end{align}As $A_{kj}$ is Hurwitz stable, i.e., ${\rm Re}(\cb{\beta}_m) < 0 ,\forall m \in \underline{\mathcal{V}}$, and $\delta_1\cb{q}_1=\frac{1}{N}\mathbb{1}^N$, it follows from \eqref{eq:impacts of variation part on secondary control input} that
\begin{align}{\label{eq:Inifity of various part}}
\langle\cb{\psi}_{a1}(\infty)\rangle=-\sum_{m=1}^3  \frac{k_Ia_{ij}^c\eta_ma_{m2}}{NI_{tj}^s\beta_m}.
\end{align}Meanwhile, it is noted that $(\frac{1}{\beta_m}, \cb{a}_m), \forall m\in \underline{\mathcal{V}}$ are eigenvalue eigenvector pairs of $\cb{A}_{kj}^{-1}$. Then, following $e^{A_{kj}(t-T_a)}\check{\cb{\phi}}_{i,j}(T_a)=\sum_{m=1}^3\eta_me^{\beta_m(t-T_a)}\cb{a}_m$, we obtain
\begin{align}\label{eq:invert of Akj}
e^{A_{kj}^{-1}}(t-T_a)\check{\cb{\phi}}_{i,j}(T_a)=\sum_{m=1}^3\eta_me^{\frac{1}{\beta_m}(t-T_a)}\cb{a}_m.
\end{align}
Differentiating both sides of equation \eqref{eq:invert of Akj} and letting $t=T_a$, we have
\begin{align}\label{eq:Intermediate output}
A_{kj}^{-1}\check{\cb{\phi}}_{i,j}(T_a)=\sum_{m=1}^3\frac{\eta_m}{\beta_m}\cb{a}_m.
\end{align}Substituting \eqref{eq:Intermediate output} into \eqref{eq:Inifity of various part}, we obtain
\begin{align}\label{eq:Inifity of various part2}
\langle\cb{\psi}_{a1}(\infty)\rangle=- \frac{k_Ia_{ij}^c}{NI_{tj}^s}\cb{k}^{\rm T}A_{kj}^{-1}\check{\cb{\phi}}_{i,j}(T_a).
\end{align}Integrating \eqref{eq:Inifity of constant part} with \eqref{eq:Inifity of various part2}, the total attack impact is
\begin{align*}
\langle\cb{\psi}_a(\infty)\rangle=-\frac{k_Ia_{ij}^c}{NI_{tj}^s}\cb{k}^{\rm T}A_{kj}^{-1}(\bar{E}_j\bar{\cb{d}}_{ij}^a(t-T_a)+\check{\cb{\phi}}_{i,j}(T_a)).
\end{align*}

Obviously, if $\cb{k}^{\rm T}A_{kj}^{-1}\bar{E}_j\bar{\cb{d}}_{ij}^a$ is not equal to zero, then $\langle \cb{\psi}_a(t) \rangle$ will diverge with $t$. Thus, $\langle \cb{\psi}(t) \rangle$ will also diverge due to $\langle\cb{\psi}(t)\rangle=\langle\tilde{\cb{\psi}}(t)\rangle+\langle\cb{\psi}_a(t)\rangle$, where the healthy component of the secondary control input vector $\langle \tilde{\cb{\psi}}(t) \rangle$ decays exponentially to zero as indicated by \eqref{eq:21}. Thus, the APV $\langle\cb{v}(t)\rangle$ will grow like ramp signals, i.e., {\em voltage balancing is not achieved}. Moreover, with $\cb{k}^{\rm T}A_{kj}^{-1}\bar{E}_j\bar{\cb{d}}_{ij}^a \ne \mathbb{0}^3$, we have $\langle C_{ij}^a\cb{k}^{\rm T}\cb{\phi}_{i,j}(\infty)\cb{l}_i \rangle = -\frac{k_Ia_{ij}^c}{NI_{tj}^s}\cb{k}^{\rm T}A_{kj}^{-1}\bar{E}_j\bar{\cb{d}}_{ij}^a \ne 0$, indicating that the equilibrium of equation \eqref{eq:intergrated Secondary control input} is not achieved, i.e.,
\begin{align}
\dot{\cb{\psi}}(t)=\tilde{L}D\cb{i}_t(\infty)-\frac{k_Ia_{ij}^c}{I_{tj}^s}\cb{k}^{\rm T}\cb{\phi}_{i,j}(\infty)\cb{l}_{i} \ne \mathbb{0}^N,
\end{align}Accordingly, given \eqref{eq:secondary control input}, {\em current sharing is not achieved either}. The proof is completed.

\subsection{Proof of Theorem \ref{Theorem:Cooperative attack imapcts} }\label{appendix:Theorem 3}
Given the linear differential equation \eqref{eq:dynamcis of the seconadry control input under attacks}, the impact caused by compromising multi communication links are the sum of the impact that would have been caused by compromising each communication link individually. Then, when cooperative ZTS attack vectors \eqref{Attack vector of the ZTS attack} satisfying \eqref{eq:cooperative attacks conditions} are injected multi communication links $\tilde{\mathcal{E}}_{c}$, simultaneously, we obtain 
\begin{align}\label{eq:28}
\langle \cb{\psi}_a(\infty) \rangle=-\sum_{(i,j) \in \tilde{\mathcal{E}}_{c}} \frac{k_Ia_{ij}^c}{NI_{tj}^s}\cb{k}A_{kj}^{-1}(\bar{E}_j\bar{\cb{d}}_{ij}^a(t-T_a)+\check{\cb{\phi}}_{i,j}(T_a)).
\end{align}Substituting \eqref{eq:cooperative attacks conditions} into \eqref{eq:28}, we have
\begin{align}\label{eq:Cooperative attack impacts 2}
\langle \cb{\psi}_a(\infty) \rangle=-\sum_{(i,j) \in \tilde{\mathcal{E}}_{c}} \frac{k_Ia_{ij}^c}{NI_{tj}^s}\cb{k}A_{kj}^{-1}\check{\cb{\phi}}_{i,j}(T_a).
\end{align}
Similar as the proof of Theorem \ref{theorem:attack impacts}, {\em voltage balancing will not be achieved if $\sum_{(i,j) \in \tilde{\mathcal{E}}_{c}} \frac{k_Ia_{ij}^c}{I_{tj}^s}\cb{k}^{\rm T}A_{kj}^{-1} \check{\cb{\phi}}_{i,j}(T_a) \ne 0$, i.e., $\langle \cb{\psi}_a(\infty)\rangle \ne 0$}. Moreover, given \eqref{eq:cooperative attacks conditions}, we obtain $\langle \sum_{(i,j) \in \tilde{\mathcal{E}}_{c}}\frac{k_Ia_{ij}^c}{I_{tj}^s}\cb{k}^{\rm T}A_{kj}^{-1}\bar{E}_j\bar{\cb{d}}_{ij}^a\cb{l_{i}} \rangle=0$, indicating that the equilibrium of $\cb{\psi}(t)$ will be achieved as $\langle \tilde{L}D\cb{i}_t(\infty) \rangle = 0$, i.e., 
\begin{align*}
\dot{\cb{\psi}}(t)=\tilde{L}D\cb{i}_t(\infty)-\sum_{(i,j) \in \tilde{\mathcal{E}}_{c}}\frac{k_Ia_{ij}^c}{I_{tj}^s}\cb{k}^{\rm T}A_{kj}^{-1}\bar{E}_j\bar{\cb{d}}_{ij}^a\cb{l}_{i} = \mathbb{0}^N.
\end{align*}To achieve current sharing, i.e., $\tilde{L}D\cb{i}_t(\infty)=\mathbb{0}^N$, it is necessary to make $\sum_{(i,j) \in \tilde{\mathcal{E}}_{c}}\frac{k_Ia_{ij}^c}{I_{tj}^s}\cb{k}^{\rm T}A_{kj}^{-1}\bar{E}_j\bar{\cb{d}}_{ij}^a\cb{l}_{i} = \mathbb{0}^N$. On the contrary, if $\sum_{(i,j) \in \tilde{\mathcal{E}}_{c}}\frac{k_Ia_{ij}^c}{I_{tj}^s}\cb{k}^{\rm T}A_{kj}^{-1}\bar{E}_j\bar{\cb{d}}_{ij}^a\cb{l}_{i} \ne \mathbb{0}^N$, then current sharing cannot be achieved. The proof is completed.

\subsection{Proof of Theorem \ref{propos: detectability for ZTS and ZTS attacks}}\label{appendix:proof of Theorem 4}
Since voltage balancing can always be achieved in the absence of attacks, i.e., $\langle \tilde{\cb{\psi}}(t) \rangle=0$, we have 
\begin{align}\label{eq:46}
\langle \cb{v}(\infty) \rangle=V_{ref}+\langle \cb{\psi}_a(\infty) \rangle.
\end{align}Moreover, under Assumption \ref{Ass:consatnt faked unknown inputs}, the DAC estimators \eqref{eq:DAC estimator} can always achieve RAC. Accordingly, substituting \eqref{eq:46} and \eqref{eq:robust average consensus} into \eqref{eq:estimated average PCC voltage deviation}, we obtain
\begin{align}\label{eq:55}
\widehat{V}_i(\infty)=\langle \cb{\psi}_a(\infty) \rangle.
\end{align}According to Theorems \ref{theorem:attack impacts}-\ref{Theorem:Cooperative attack imapcts}, under ZTS attacks with constant $\bar{\cb{d}}_{ij}^a,\forall (i,j)\in \tilde{\mathcal{E}}_c$, PCC voltages will either converge to stable values or grow like ramp signals. It is obvious that, if $\langle\cb{\psi}_a(\infty) \rangle$ is ramp-growing, then the detection indicator $\mathcal{d}_i(t)$ will also keep growing due to \eqref{eq:55}, indicating that \eqref{eq:normal condition} will be eventually violated. Otherwise, if $\langle\cb{\psi}_a(\infty) \rangle$ is a constant, then with \eqref{eq:detectability condition}, we have
\begin{align*}
\mathcal{d}_i(\infty)=T\langle \cb{\psi}_a(\infty) \rangle>\bar{\mathcal{d}}_i,
\end{align*}which means that \eqref{eq:normal condition} is also violated. The state follows.

\subsection{Proof of Theorem \ref{propos:effectiveness of the counteraction}}\label{appendix:proof of theorem 5}
First, we consider the cooperative ZTS attacks satisfying \eqref{eq:cooperative attacks conditions}. According to the result in Theorem \ref{Theorem:Cooperative attack imapcts}, cooperative ZTS attacks with constant $\bar{\cb{d}}_{ij}^a,\forall (i,j)\in \tilde{\mathcal{E}}_c$ will eventually cause constant APVD and let $\langle \cb{\psi}_a(\infty) \rangle=-\sum_{(i,j) \in \tilde{\cb{\varepsilon}}_{a}}\frac{k_Ia_{ij}^c}{NI_{tj}^s}\cb{k}^{\rm T}A_{kj}^{-2}\bar{E}_j\bar{\cb{d}}_{ij}^a=\bar{Q}_c^b$. After activating the impact counteraction strategy \eqref{eq:impact counteraction}, the APV is obtained as 
\begin{align}\label{eq:49}
\langle\cb{v}(\infty)\rangle=V_{ref}+\bar{Q}_c^b+\langle\cb{c}(\infty)\rangle,
\end{align}where $\cb{c}(t)$ collects the compensation values $C_i(t),\forall i \in \mathcal{V}$. Since the DAC estimators \eqref{eq:DAC estimator} can achieve RAC, we have
\begin{align}\label{eq:50}
\widehat{V}_i(\infty)=\langle\cb{v}(\infty)\rangle.
\end{align}Substituting equations \eqref{eq:49} and \eqref{eq:50} into \eqref{eq:estimated average PCC voltage deviation}, we obtain
\begin{align}\label{eq:51}
\widehat{V}_i^{err}(\infty)=- \bar{Q}_c^b-\langle\cb{c}(\infty)\rangle.
\end{align}Integrating the PI-based compensator \eqref{eq:Compensator} with \eqref{eq:51}, it is obvious that the equilibrium of \eqref{eq:Compensator} will be achieved with $\widehat{V}_i^{err}(\infty)=0$, under which the voltage balancing is achieved.

Then, we consider the non-cooperative ZTS attacks where \eqref{eq:cooperative attacks conditions} is not satisfied. Based on the result in Theorem \ref{theorem:attack impacts}, non-cooperative ZTS attacks with constant $\bar{\cb{d}}_{ij}^a,\forall (i,j)\in \tilde{\mathcal{E}}_c$ will cause ramp-growing APVD and let 
{\begin{align*}
\langle \cb{\psi}_a(\infty) \rangle&=\sum_{(i,j) \in \tilde{\mathcal{E}}_{c}}-\frac{k_Ia_{ij}^c}{NI_{tj}^s}\cb{k}^{\rm T}A_{kj}^{-1}(\bar{E}_j\bar{\cb{d}}_{ij}^a(t-T_a)+A_{kj}^{-1}\bar{E}_j\bar{\cb{d}}_{ij}^a)=\bar{Q}_n^kt+\bar{Q}_n^b.
\end{align*}}Similar to \eqref{eq:51}, we have
\begin{align*}
\widehat{V}_i^{err}(\infty)=-\bar{Q}_n^kt-\bar{Q}_n^b-\langle\cb{c}(\infty)\rangle.
\end{align*}To track the ramp-growing signal $\bar{Q}_n^kt+\bar{Q}_n^b$ with the PI-based compensator \eqref{eq:Compensator}, there should be a nonzero tracking error such that 
\begin{align*}
k_{ci}\widehat{V}_i^{err}(\infty)=\bar{Q}_n^k.
\end{align*}

Thus, the APV after compensation can be written as \eqref{eq:average PCC voltage after compensation}, and the proof is completed.

\subsection{Proof of Lemma \ref{propos:design of h(s) and g(s)}}\label{appendix:proof of lemma 1}
According to Theorem 2 in \cite{bai2010robust}, we obtain that, under undirected connected graph $\mathcal{G}_c$ and the input signals satisfying $V_i(s)=\frac{c_i(s)}{d(s)}, \forall i \in \mathcal{V}$, the DAC estimators \eqref{eq:DAC estimator} can achieve RAC if the corresponding transfer functions $h(s)=\frac{n_h(s)}{d_h(s)},g(s)=\frac{n_g(s)}{d_g(s)}$ satisfy

\begin{enumerate}
  \item[a)] there exists the polynomial $p_h(s)$ such that $n_h(s)-d_h(s)=p_h(s)d(s)$ and $h(s)$ is stable;
  \item[b)] there exists the polynomial $p_g(s)$ such that $d_g(s)=p_g(s)d(s)$;
  \item[c)] all roots of $d_h(s)d_g(s)+n_h(s)n_g(s)\gamma^2\lambda_i^2=0, \forall i \in \{2,\cdots, N\}$ lie in the open left half-plane, where $0=\lambda_1<\lambda_2\le\cdots\le\lambda_N$ are eigenvalues of the Laplacian matrix $L$.
\end{enumerate}Thus, under Assumption \ref{Ass:graph} and the input signals satisfying $V_i(s)=\frac{c_i^c(s)}{s}+\frac{c_i^r(s)}{s^2}, \forall i \in \mathcal{V}$, we can verify whether DAC estimators \eqref{eq:DAC estimator} can achieve RAC through the conditions a)-c). First, it is intuitive that conditions a) and b) are satisfied when $h(s)$ and $g(s)$ are set as \eqref{eq:h(s) and g(s)} and $d(s)=s$ or $s^2$. Next, it is shown that condition c) is also satisfied for arbitrary $\lambda_i >0, \forall i \in \{2,\cdots,N\}$ and $\gamma$. Integrated with \eqref{eq:h(s) and g(s)}, the equation $d_h(s)d_g(s)+n_h(s)n_g(s)\gamma^2\lambda_i^2=0$ is transformed to 
\begin{align*}
(s+a)(s^2(s+a)+(2as+a^2)\gamma^2\lambda_i^2)=0,
\end{align*}under which we obtain that $s=-a$ is the negative real root. The remaining roots are determined by
\begin{align}\label{eq:35}
f(s)=s^3+as^2+2a\gamma^2\lambda_i^2s+a^2\gamma^2\lambda_i^2=0.
\end{align}According to the Cardano's formula, the nature of roots corresponding to \eqref{eq:35} can be observed by utilizing the following discriminant without directly computing them, i.e.,
\begin{align*}
\Delta&=(\frac{1}{27}a^3+\frac{1}{6}a^2\gamma^2\lambda_i^2)^2+(\frac{2}{3}a\gamma^2\lambda_i^2-\frac{1}{9}a^2)^3=\frac{32a^3\gamma^2\lambda_i^2}{108}(\gamma^4\lambda_i^4-\frac{13}{32}a\gamma^2\lambda_i^2+\frac{1}{8}a^2).
\end{align*}With $a>0$, we have
\begin{align*}
\Delta>\frac{32a^3\gamma^2\lambda_i^2}{108}(\gamma^2\lambda_i^2-\frac{13}{64}a)^2\ge0,
\end{align*}indicating that the cubic equation \eqref{eq:35} has one real root $s_1$ and two complex conjugate roots $s_2,s_3$, which satisfy
\begin{subequations}
\begin{alignat}{2}
&s_1+s_2+s_3=-a, \label{eq:36_1} \\
&s_1s_2s_3=-a^2\lambda_i. \label{eq:36_2}
\end{alignat}
\end{subequations}Given \eqref{eq:36_2} and $s_2s_3>0$, we derive $s_1<0$. Moreover, it follows from \eqref{eq:36_1} that the complex roots $s_2,s_3$ lie in the open left half-plane if $0>s_1>-a$. The differential of the cubic function \eqref{eq:35} corresponds to
\begin{align*}
\dot{f}(s)=3s^2+2as+2a\gamma^2\lambda_i^2=3(s+\frac{1}{3}a)^2+\frac{a}{3}(6\gamma^2\lambda_i^2-a),
\end{align*}which is non-negative if $a\le6\gamma^2\lambda_i^2$, i.e., $f(s)$ is monotonically increasing. Hence, with $f(-a)=-a^2\lambda_i<0$, we infer that $s_1>-a$. Otherwise, if $a>6\gamma^2\lambda_i^2$, then $\dot{f}(s)$ can be negative and let $\tilde{s}_1<\tilde{s}_2<0$ be the corresponding roots such that $\dot{f}(\tilde{s}_1)=\dot{f}(\tilde{s}_2)=0$. Since $\tilde{s}_1+\tilde{s}_2 =- \frac{2a}{3}$, we obtain that $\tilde{s}_1>-a$, indicating that $f(s)$ is monotonically increasing when $s\in(-\infty,-a]$. Hence, it follows from $f(-a)<0$ that $f(s)<0,\forall s\in(-\infty,-a]$, and then $s_1>-a$ is guaranteed. Therefore, regardless of $\gamma,\lambda_i$, conditions a)-c) are satisfied if $h(s),g(s)$ are set according to \eqref{eq:h(s) and g(s)} and $a>0$. The proof is completed.

\subsection{Proof of Lemma \ref{lemma:UIOs for DAC estimators}}\label{appendix:proof for lemma 2}
It follows from \eqref{eq:minimal realizations of LTIs} and \eqref{eq:UIO for DAC estimators} that
\begin{subequations}
\begin{alignat*}{2}
{\rm rank}({\rm I}^2B_1)={\rm rank}(B_1)=1,\\ 
{\rm rank}({\rm I}^2B_2)={\rm rank}(B_2)=1,
\end{alignat*}
\end{subequations}and matrices
\begin{align*}
\left[ 
\begin{array}{cc}
s{\rm I}^2-A_1 & B_1 \\
{\rm I}^2 & 0
\end{array} 
\right],
\left[ 
\begin{array}{cc}
s{\rm I}^2-A_2 & B_2 \\
{\rm I}^2 & 0
\end{array} 
\right] 
\end{align*}
both have full column rank $\forall s \in \mathbb{C}$. Hence, based on \eqref{eq:DAC estimator}, one can construct UIOs \eqref{eq:UIO for DAC estimators} in DGU $i$ such that, in the absence of attacks, detection residuals $\cb{r}_{i,j1}^{v}(t), \cb{r}_{i,j2}^{\eta}(t)$ can be written as \eqref{eq:healthy component of residual for DAC}. Since $F_{j}^{v},F_{j}^{\eta}$ are designed to be Hurwitz stable, $\cb{r}_{i,j1}^{v}(t), \cb{r}_{i,j2}^{\eta}(t)$ will both decay exponentially to zero. Moreover, with ${\rm rank}({\rm I}^2)>{\rm rank}(B_1)={\rm rank}(B_1)$, the UIO-based detectors \eqref{eq:UIO for DAC estimators} can detect FDI attacks on communicated DAC related information between DGUs.

\subsection{Parameters of DGUs}\label{appendix:electrical_parameyters}
\begin{table}[h]
\centering 
\caption{Electrical Parameters of DGUs}
\begin{tabular}{p{1cm}<{\centering}p{0.8cm}<{\centering}p{0.8cm}<{\centering}p{0.8cm}<{\centering}p{1cm}<{\centering}p{0.8cm}<{\centering}p{0.8cm}<{\centering}p{0.8cm}<{\centering}}
\toprule[1.2pt]
\multicolumn{8}{c}{\textbf{Converter Parameters}} \\ \hline
DGU & $R_t({\rm \Omega})$ & $L_t({\rm mH})$ & $C_t({\rm mF})$ & DGU & $R_t({\rm \Omega})$ & $L_t({\rm mH})$ & $C_t({\rm mF})$\\ \hline
DGU$1$ & $0.2$ & $1.8$ & $2.2$ & DGU$2$ & $0.4$ & $2.0$ & $1.7$ \\ 
DGU$3$ & $0.3$ & $2.2$ & $1.9$ & DGU$4$ & $0.6$ & $2.5$ & $2.4$ \\ 
DGU$5$ & $0.5$ & $3.0$ & $2.7$ & DGU$6$ & $0.4$ & $1.6$ & $3.0$ \\ 
DGU$7$ & $0.2$ & $1.4$ & $2.1$ & DGU$8$ & $0.4$ & $1.2$ & $1.6$ \\ 
\toprule[1.2pt]
\multicolumn{8}{c}{\textbf{Power Line Parameters}} \\ \hline
$(i,j)$ & $R_{ij}({\rm \Omega}) $ & $(i,j)$ & $R_{ij}({\rm \Omega})$ & $(i,j)$ & $R_{ij}({\rm \Omega}) $ & $(i,j)$ & $R_{ij}({\rm \Omega})$\\
$(1,2)$ & $0.05$ & $(1,6)$ & $0.10$ & $(2,3)$ & $0.07$ & $(3,4)$ & $0.09$ \\
$(3,8)$ & $0.12$ & $(4,5)$ & $0.14$ & $(4,8)$ & $0.20$ & $(5,6)$ & $0.25$ \\
$(5,7)$ & $0.06$ & $(6,7)$ & $0.10$ & $(1,7)$ & $0.15$ & $(8,7)$ & $0.17$ \\
\toprule[1.2pt]
\end{tabular}
\end{table}

\begin{table}[h]
\centering 
\caption{Current Loads and Rated Currents in DGUs}
\begin{tabular}{p{1cm}<{\centering}p{0.8cm}<{\centering}p{0.8cm}<{\centering}p{0.8cm}<{\centering}p{0.8cm}<{\centering}p{0.8cm}<{\centering}}
\toprule[1.2pt]
\multicolumn{6}{c}{\textbf{Current loads and rated currents}} \\ \hline
DGU & $I_{Li}({\rm A})$ & $I_{ti}^s({\rm A})$ & DGU & $I_{Li}({\rm A})$ & $L_{ti}^s({\rm A})$\\ \hline
DGU$1$ & $10$ & $20$ & DGU$2$ & $10$ & $20$ \\ 
DGU$3$ & $15.75$ & $30$ & DGU$4$ & $9$ & $20$ \\ 
DGU$5$ & $14$ & $20$ & DGU$6$ & $21$ & $30$ \\ 
DGU$7$ & $16.25$ & $25$ & DGU$8$ & $18.75$ & $25$ \\ 
\toprule[1.2pt]
\end{tabular}
\end{table}

\subsection{The impact of perturbing $a_{ij}^{cd}$ and $a$}
In this subsection, we show the impact of perturbing $a_{ij}^{cd},\forall (i,j)\in\mathcal{E}$ or $a$ through two independent cases, where $\hat{V}_i^{err}(t)$ and $\mathcal{d}_i(t)$ are depicted and the quantities after perturbation are denoted by $\tilde{a}_{ij}^{cd}$ and $\tilde{a}$. The perturbation is introduced at $t=6$s. As shown in Figs. \ref{fig:PerturbationImpacts_EdgeWeights}-\ref{fig:PerturbationImpacts_Parametera}, under the perturbation, the estimated APVDs will finally converge to zero, and the fluctuation of the estimated APVDs can be tolerated by the threshold $\bar{\mathcal{d}}_i=0.0325$ when $\tilde{a}_{ij}^{cd}\in [a_{ij}^{cd},20a_{ij}^{cd}]$ and $\tilde{a} \in [a,1.1a]$. Hence, the impact of perturbing $a_{ij}^{cd},\forall (i,j)\in\mathcal{E}$ and $a$ can be limited such that the performance in detecting ZTS attacks will not be degraded.

\begin{figure}[htbp]
\includegraphics[scale=0.3]{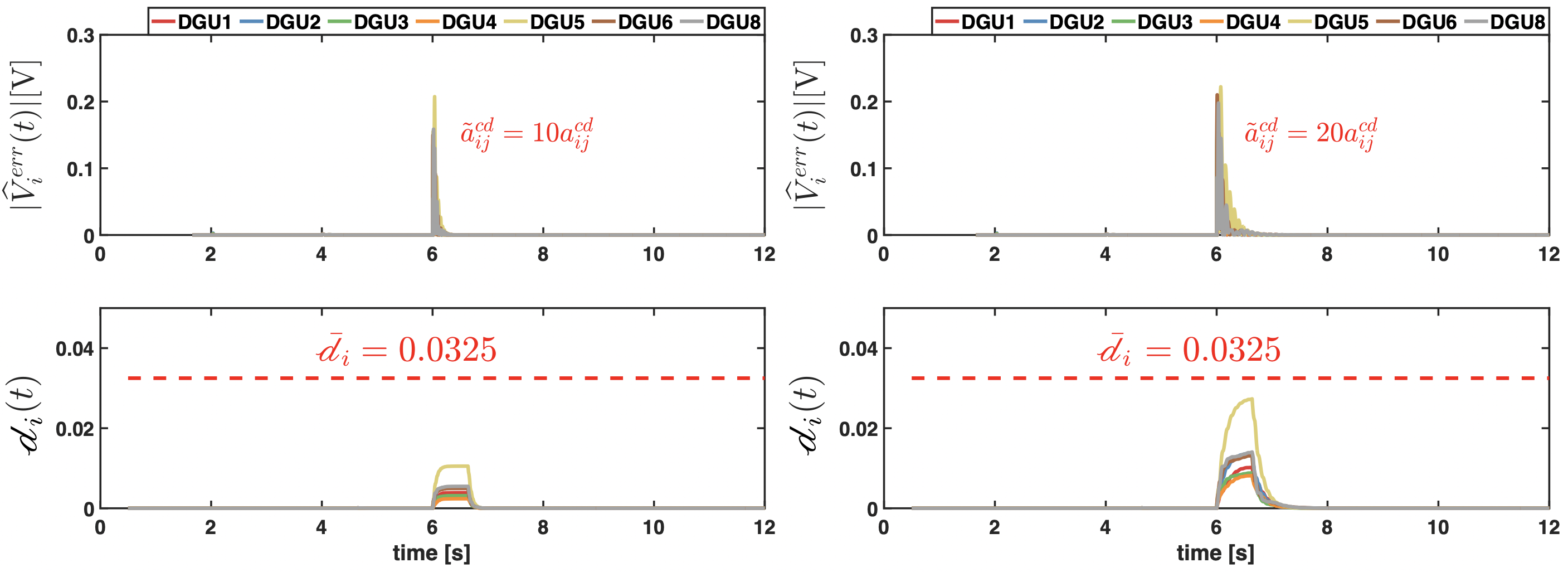}
\centering
\caption{This figure shows the impact of perturbing $a_{ij}^{cd},\forall (i,j)\in\mathcal{E}_c$.} \label{fig:PerturbationImpacts_EdgeWeights}
\end{figure}

\begin{figure}[htbp]
\centering
\includegraphics[scale=0.3]{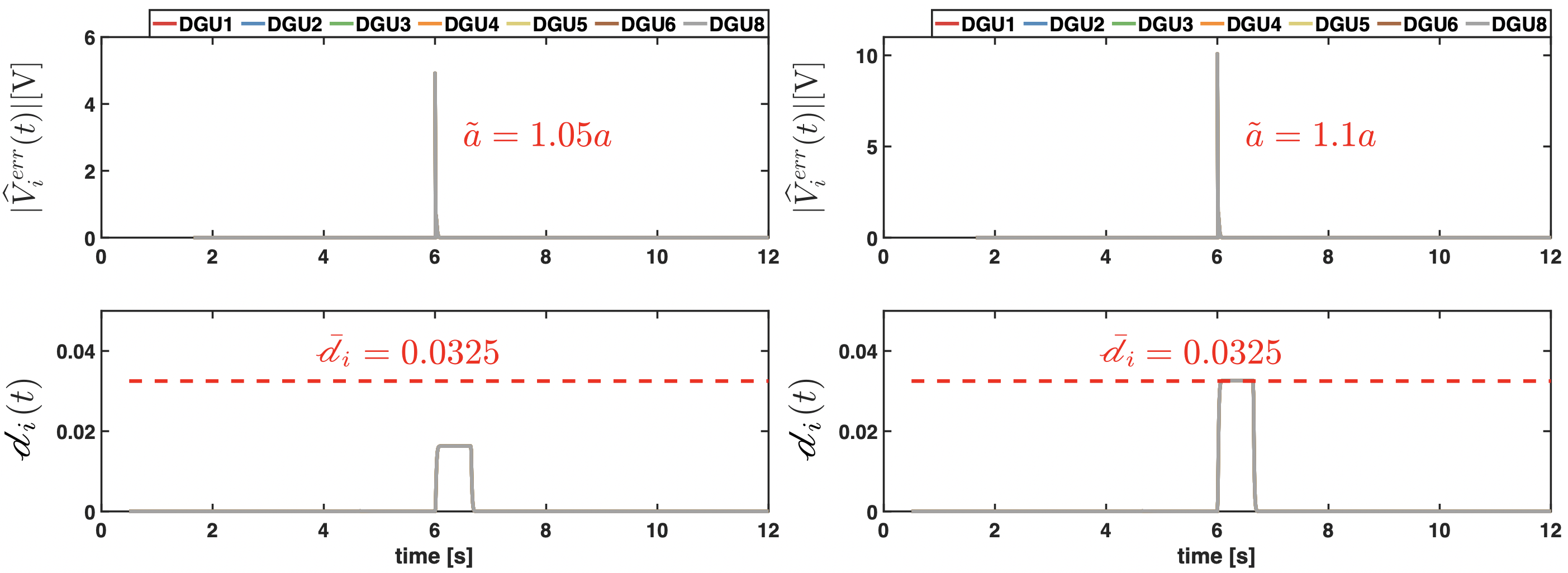}
\caption{This figure shows the impact of perturbing $a$.} \label{fig:PerturbationImpacts_Parametera}
\end{figure}

\subsection{The impact of the false alarm caused by daily operations}
In this subsection, we show the impact of the false alarm in the DCmG, which is caused by decreasing all DGU loads by $40$\% simultaneously at $t=6$s. According to Fig. \ref{fig:systemstates under false alarm}, under the false alarm, voltage balancing can be still achieved, which validates the statement in Remark \ref{Remark:impact mitigation}. Moreover, compared with the case without compensation, voltage balancing can be recovered with a quicker rate under the impact mitigation strategy \eqref{eq:impact counteraction}.}

\begin{figure}[htbp]
\includegraphics[scale=0.3]{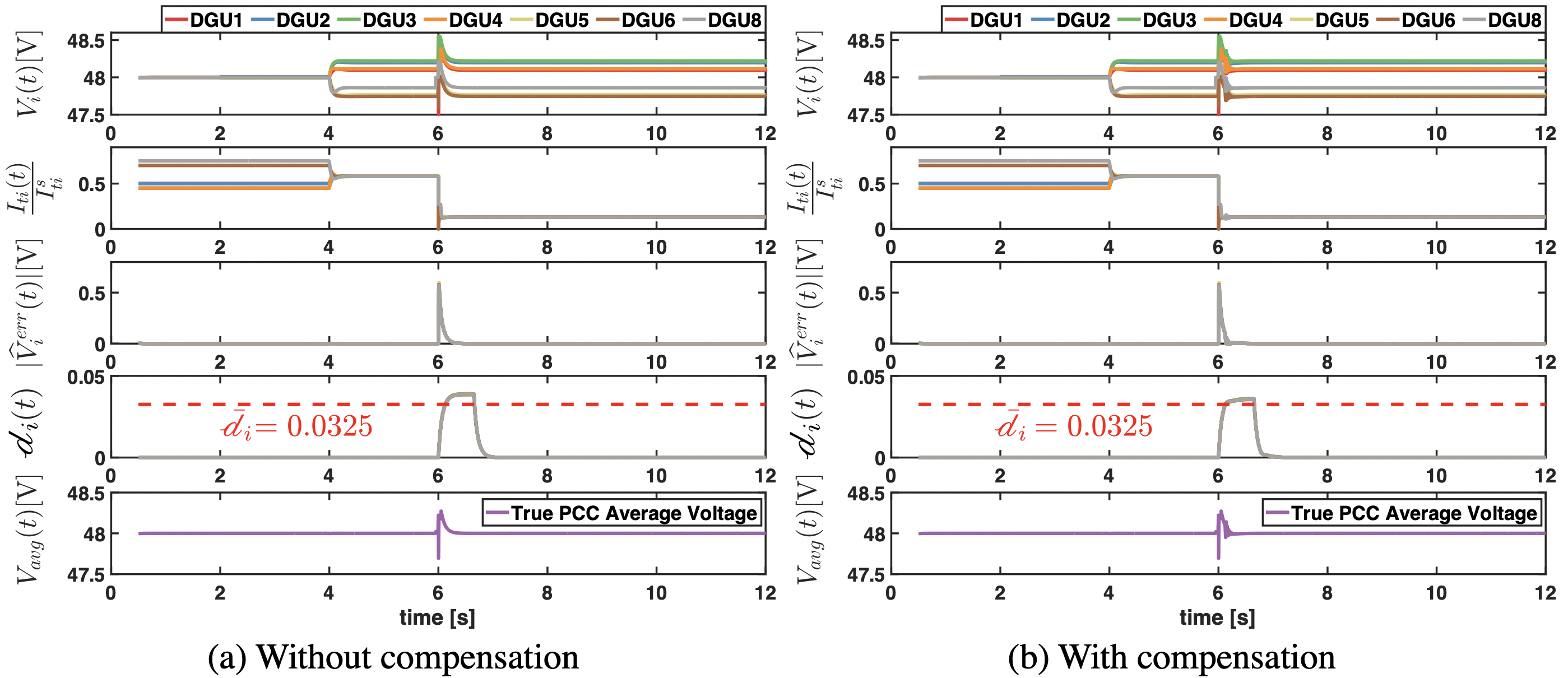}
\centering
\caption{This figure shows PCC voltages, output currents, and countermeasure related variables under the false alarm without compensation ($k_{cp}=k_{ci}=0$) and with compensation ($k_{cp}=1,k_{ci}=20$).} \label{fig:systemstates under false alarm}
\end{figure}

\end{document}